\documentclass[aps,jcp,amsmath,amssymb,showpacs,letterpaper,twocolumn,superscriptaddress]{revtex4}%
\usepackage{hyperref}
\usepackage{bm}
\usepackage{amsmath}
\usepackage{graphicx}%
\usepackage{amsfonts}%
\usepackage{amssymb}
\usepackage{color}
\usepackage{setspace}
\usepackage{bigints}
\usepackage{xfrac} 
\usepackage{multirow}
\usepackage{booktabs}
\usepackage{lipsum}

\usepackage{afterpage}

\renewcommand*{\today}{April 10, 2022}

\begin{document}

\title{The interplay of cation/anion and monovalent/divalent selectivity in negatively charged nanopores: local charge inversion and anion leakage} 

\author{Eszter Lakics}
\affiliation{Center for Natural Sciences, University of Pannonia, Egyetem u.\ 10, Veszpr\'{e}m, 8200, Hungary}
\author{M\'onika Valisk\'o}
\affiliation{Center for Natural Sciences, University of Pannonia, Egyetem u.\ 10, Veszpr\'{e}m, 8200, Hungary}
\author{Dirk Gillespie}
\affiliation{Department of Physiology and Biophysics, Rush University Medical Center, Chicago, Illinois 60612, USA}
\author{Dezs\H{o} Boda}
\email{boda.dezso@mk.uni-pannon.hu}
\affiliation{Center for Natural Sciences, University of Pannonia, Egyetem u.\ 10, Veszpr\'{e}m, 8200, Hungary}
\date{\today}

\begin{abstract}
The anomalous mole fraction effect (AMFE) is widely regarded as a hallmark of calcium versus monovalent ion selectivity in negatively charged pores. 
While AMFE is well understood in highly cation-selective narrow ion channels, its microscopic origin in wide synthetic nanopores, where anions may also contribute to transport, remains less clear. 
Here, we use a reduced Nernst-Planck + Local Equilibrium Monte Carlo framework to study ionic transport in a negatively charged PET nanopore, with particular emphasis on how the modeling of surface carboxyl (COO$^{-}$) groups influences charge inversion, ionic currents, and AMFE. 
We systematically compare fixed point-charge models and explicit-particle representations of surface oxygens and identify two controlling parameters: the distance of closest approach (DCA) between ionic charges and pore charges and grid spacing that modulates localization (while keeping average surface charge constant).
By fitting pore diffusion coefficients to three experimental conductance points, we reproduce the entire experimental AMFE curve as well as anion leakage in CaCl$_2$ seen in experiments and molecular dynamics simulations. 
Remarkably, vastly different microscopic models of the surface groups yield indistinguishable device-level conductance curves when the DCA is matched, despite substantial differences in local Ca$^{2+}$ concentration profiles. 
Our results demonstrate that AMFE in wide nanopores is governed by a delicate interplay between charge inversion, anion leakage, and ionic mobility, underlying that in wide pores monovalent vs.\ divalent cation selectivity is modulated by cations vs.\ anion selecivity.
\end{abstract}

\maketitle

\section{Introduction}
\label{sec:intro}

The anomalous mole fraction effect (AMFE) is widely regarded as a defining signature of calcium (divalent) versus monovalent ion selectivity in negatively charged pores, including both biological ion channels~\cite{almers_jp_1984a,almers_jp_1984b,gillespie_jpcb_2005,gillespie_bj_2008_energetics,gillespie_giri_bj_2009} and synthetic nanopores~\cite{gillespie_bj_2008_nanopore}. 
Because these pores are overall cation-selective, the AMFE indicates a ``preferential selectivity'' between cations, defining which is more preferred~\cite{gillespie_giri_bj_2009}.
In AMFE experiments, the mole fraction of Ca$^{2+}$ in a mixed electrolyte is gradually increased while the total ionic current through the pore is monitored as a function of composition. 
The effect is termed anomalous because the resulting current-composition curve is non-monotonic, exhibiting a pronounced minimum at an intermediate Ca$^{2+}$ mole fraction. 
The lower the Ca$^{2+}$ mole fraction at which this minimum occurs, the more strongly Ca$^{2+}$-selective the pore is considered to be.

AMFE was first reported for the L-type calcium channel~\cite{almers_jp_1984a,almers_jp_1984b}, later for the ryanodine receptor (RyR) calcium release channel~\cite{gillespie_jpcb_2005,gillespie_bj_2008_energetics,gillespie_giri_bj_2009}, and subsequently for a synthetic biconical polyethylene terephthalate (PET) nanopore whose diameter is much larger than that of typical ion channels~\cite{gillespie_bj_2008_nanopore} (see \href{https://ars.els-cdn.com/content/image/1-s2.0-S0167732222022541-gr1_lrg.jpg}{Fig.~1}). 
Our earlier modeling efforts~\cite{gillespie_bj_2008_ca,boda_jgp_2009,malasics_bba_2010_trivalent,boda_jcp_2011_analyze,boda_jcp_2013_solvation,fabian_jml_2022} replaced the original single-file interpretation~\cite{Hille} with a more general physical picture: the current minimum arises because monovalent cations are expelled from the pore by Ca$^{2+}$ (depressing monovalent current), while Ca$^{2+}$ itself carries little current at low mole fraction due to its low bulk concentration.

This explanation is particularly clear for narrow biological ion channels, where anions (Cl$^{-}$) are effectively excluded and the competition occurs solely between cations in a highly charged environment that strongly prefers divalents. 
In contrast, the situation is more complex in wide nanopores (radius $\sim 2.7$ nm at the narrowest part , versus $<0.3$ nm for ion channels), where anion leakage occurs and can significantly contribute to the total current.

In this work, we advance this picture by proposing that the nanopore not only allows anion leakage in pure CaCl$_2$ solutions, but that the Cl$^{-}$ current may even exceed the Ca$^{2+}$ current. 
We demonstrate that this scenario is consistent with experimental observations~\cite{he_jacs_2009}, recent molecular dynamics simulations~\cite{shabbir_jcp_2026}, and the experimentally measured AMFE curves of Gillespie et al.~\cite{gillespie_bj_2008_nanopore}.

The reduced model employed here closely follows that used in our previous work~\cite{fabian_jml_2022}, but is extended with a more detailed representation of the deprotonated carboxyl (COO$^{-}$) groups decorating the PET nanopore wall. 
Proper modeling of these charged surface groups is essential because the central mechanism underlying AMFE in nanopores is charge inversion: Ca$^{2+}$ ions adsorb strongly to the carboxyl groups and locally overcompensate the surface charge~\cite{boda_entropy_2020}. 

In the context here, charge inversion is an interplay of all the three ionic species.
Ca$^{2+}$ ionsoutcompete K$^{+}$ ions and are tightly bound to the surface groups, which limits their mobility. 
Cl$^{-}$ ions, on the other hand, may form a layer of inverted charge where their mobility may exceed that of Ca$^{2+}$. 
K$^{+}$ ions are electrostatically repelled by the overcharged pore wall, which decreases their conductance even at low Ca$^{2+}$ mole fractions.

The magnitude of each mechanism depends on the strength of the charge inversion.
In simulations, this is in large part is controlled by how the surface groups are modeled. 
This is because charge inversion is a local phenomenon that depends sensitively on the microscopic structure of the charged surface groups. 
While our earlier work~\cite{fabian_jml_2022} already demonstrated the importance of representing surface charge in a localized manner (individual charges on a grid rather than a uniform surface charge), the present study identifies another control parameter, beyond the degree of localization, of the reduced model: the distance of closest approach (DCA) between the central charge of Ca$^{2+}$ ions and the pore charge.

Here, we systematically investigate different representations of the COO$^{-}$ groups, ranging from fixed point charges to flexible oxygen atoms modeled as charged hard spheres confined by harmonic potentials. 
This work follows our usual modeling strategy, in which complexity is introduced incrementally, progressing from simple to increasingly realistic descriptions to define which complexity is truly required in a model~\cite{boda_jpcb_2000,boda_mp_2002,boda_jcp_2006,boda_jctc_2012,boda_jcp_2013_solvation,fabian_jml_2022,shabbir_jcp_2026}.

\section{Model and method}

We employ a reduced model in which the number of degrees of freedom is deliberately small by treating selected components implicitly rather than explicitly. 
Most notably, water is modeled as a continuum medium instead of as explicit molecules, interacting with ions through effective response functions that account for dielectric screening and friction (the dielectric constant $\epsilon$ and position-dependent diffusion coefficients $D_i(\mathbf{r})$, respectively). 
The guiding principle of this approach is to explicitly retain the degrees of freedom essential for nanopore current conduction, while incorporating less critical ones implicitly through physically self-consistent averaging procedures~\cite{boda_entropy_2020}. 
Reduced models are computationally efficient, enabling simulations across a broad range of state points, and they facilitate the sampling of millimolar and sub-millimolar concentrations, which is notoriously challenging in explicit-solvent MD. 

\subsection{Model of the nanopore and electrolyte}

\begin{figure*}[t]  
	\centering
	\includegraphics[width=0.6\linewidth]{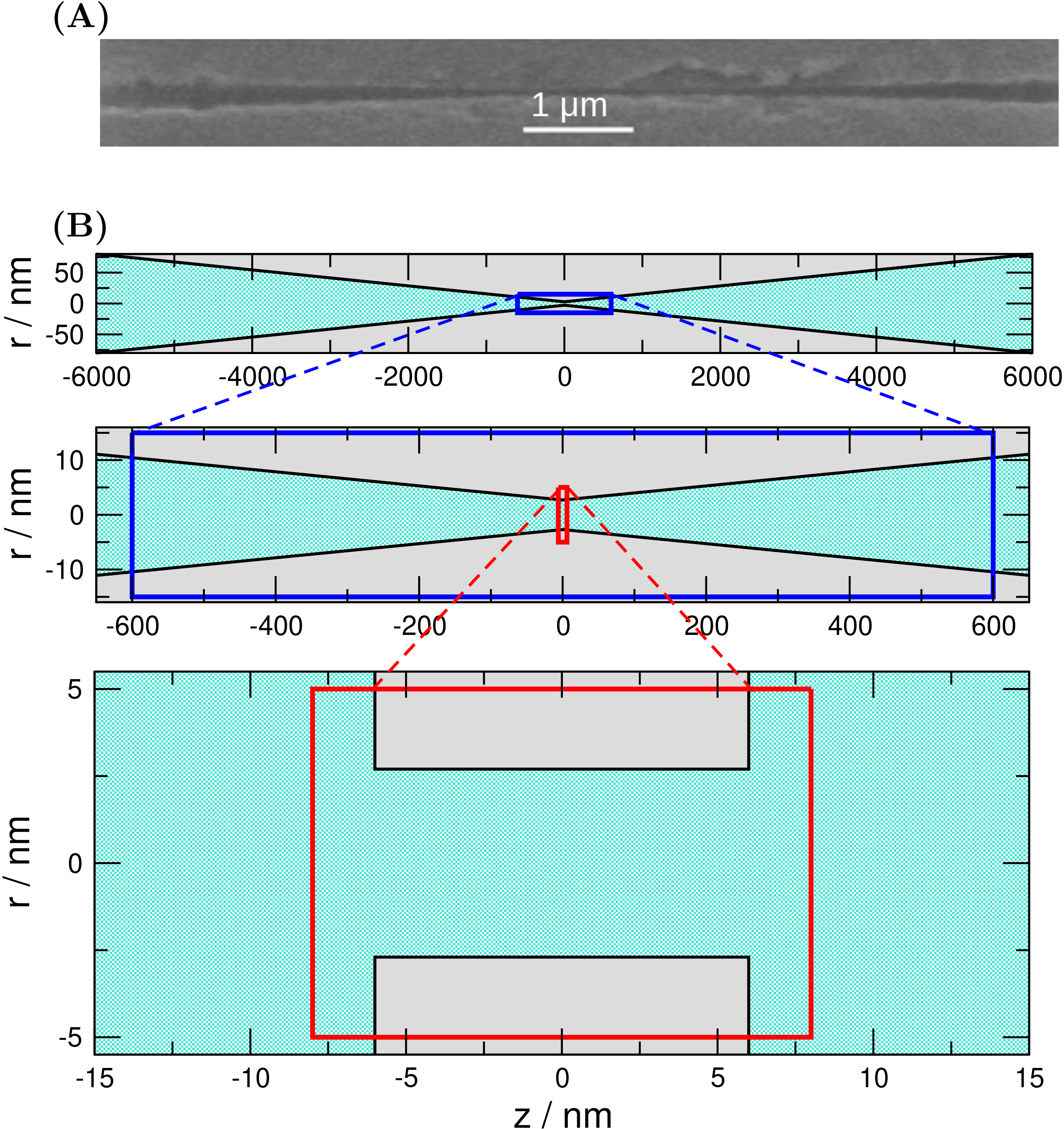} 
	\caption{\small (A) Scanning electron micrograph of a cross section of a biconical PET nanopore from the paper of Gillespie et al.~\cite{gillespie_bj_2008_nanopore}. The length of the pore (width of the PET membrane) is 12 $\mu$m. The large radii at the entrances are 79 nm, while the small radius at the center is $2.7$ nm. 
		(B) The top panel shows the whole nanopore (axes aspect ratio is not to scale). The middle panel zooms on the blue rectangle of the top panel (axes aspect ratio is not to scale). The bottom panel zooms on the red rectangle of the middle panel (aspect ratio is correct). We explicitly simulate the central $H=12$ nm long portion (red rectangle), while the effect of the rest of the pore is taken into account by tuning the diffusion coefficients in the pore, $D_{i}^{\mathrm{p}}$. In our specific simulation cell, we have two baths on the two sides of the membrane (bottom panel) with their own diffusion coefficients. }
	\label{fig1}
\end{figure*}

In the experiments of Gillespie et al.~\cite{gillespie_bj_2008_nanopore}, a long (12 $\mu$m) and narrow ($2.7-79$ nm) biconical nanopore was used.
The scanning electron micrograph of the cross section is seen in Fig.~\ref{fig1}A.
It is not feasible to handle a system of this size explicitly in a computer simulation even without reduced models.

In this work, however, we are interested in the selectivity properties of the pore, and not necessarily in the detailed conduction properties of each segment of the pore.
Therefore, we assume that the narrowest (2.7 nm radius) part of the pore (the red frame in Fig.~\ref{fig1}B) governs selectivity (its role is similar to that of the selectivity filter of an ion channel).
We also assume that it is sufficient to model this central part of the pore  explicitly, while the rest of the pore can be taken into account by adjusting the diffusion coefficients to experimental data. 
Since the whole pore is $1000$ times longer than our model pore ($12$ $\mu$m vs.\ $12$ nm), 2-3 orders of magnitude reduction of the diffusion coefficient will be needed to match the conductance values.

Note that the real nanopore shown in Fig.~\ref{fig1}A is not perfectly biconical and its walls are not perfectly smooth~\cite{YuApel2012}.
The effect of these features, however, can be included in the fitted diffusion coefficients.

The model is composed of two parts: a cylindrical nanopore (length $H=12$ nm and radius $R=2.7$ nm) embedded in a membrane and two bulk phases on either side of the membrane (bottom panel of Fig.~\ref{fig1}B).
The walls of the pore and the membrane are hard, so ion overlap with the walls is forbidden. 

We impose boundary conditions for the electrochemical potential at the boundary of the finite simulation cell, which is a cylinder containing the membrane, the pore, and the bulk regions.
In this work, we keep the concentrations of all the ionic species the same on the two sides of the membrane, while we impose different electrical potentials on the two sides to create the driving force of ionic transport (the voltage, $U$).

The ions of the electrolyte are modeled explicitly as hard spheres with point charges in their centers.
The spheres cannot overlap, either with the membrane wall or with each other. 
The interaction between ions is
\begin{equation}
u(r) =
\left\lbrace 
\begin{array}{ll}
\infty & \quad \mathrm{for} \quad r<R_{i}+R_{j} \\
 \dfrac{1}{4\pi\epsilon_{0}\epsilon} \dfrac{q_{i}q_{j}}{r} & \quad \mathrm{for} \quad r \geq R_{i}+R_{j}\\
\end{array}
\right. 
\label{eq:uij},
\end{equation}
where $R_{i}$ and $R_{j}$ are the ionic radii of species $i$ and $j$, $\epsilon_0$ is the permittivity of vacuum, $r$ is the distance between the centers of the ions, and $q_{i}$ and $q_{j}$ are the charges of the ions ($q_{i}=z_{i}e$, where $z_{i}$ is the ionic valence and $e$ is the fundamental charge).

We use the Pauling radii as in many of our previous studies for ion channels~\cite{boda_jcp_2006,gillespie_bj_2008_ca,boda_jgp_2009,malasics_bba_2010_trivalent,boda_jcp_2013_solvation,boda_arcc_2014,boda_entropy_2020}, nanopores~\cite{gillespie_bj_2008_nanopore,he_jacs_2009,valisko_jcp_2019}, and bulk solutions~\cite{vincze_jcp_2010,valisko_jcp_2014,valisko_jpcb_2015,valisko_mp_2017}.
The effect of the hydration shell on the adsorption of ions at the charged surfaces can be taken into account by an adjusted distance of closest approach of the cations to the localized charged groups. 
Experimental diffusion constants are applied in the bulk phase ($D_{i}^{\mathrm{b}}$, see Table \ref{tab}). 

\begin{table}
\centering
\caption{Pauling radii~\cite{pauling} and bulk-phase diffusion constants~\cite{haynes2016crc} of ions.} \label{tab}
\vspace{0.2cm}
\begin{tabular}{lll}
\hline
 Ion & $R_{i}$ / nm & $D_{i}^{\mathrm{b}}$ / m$^{2}$s$^{-1}$ \\ \hline
  Li$^{+}$ & $0.06$ & $1.029\times 10^{-9}$ \\
  Na$^{+}$ & $0.095$ & $1.334\times 10^{-9}$ \\
 K$^{+}$ & $0.133$ & $1.96\times 10^{-9}$ \\
 Ca$^{2+}$ & $0.099$ & $0.792\times 10^{-9}$ \\
 Cl$^{-}$ & $0.181$ & $2.032\times 10^{-9}$ \\ \hline
\end{tabular}
\end{table}

\subsection{Models for the charged groups on the pore surface}
\label{sec:pore_charges}

\begin{figure*}[t]  
	\centering
	\includegraphics[width=\linewidth]{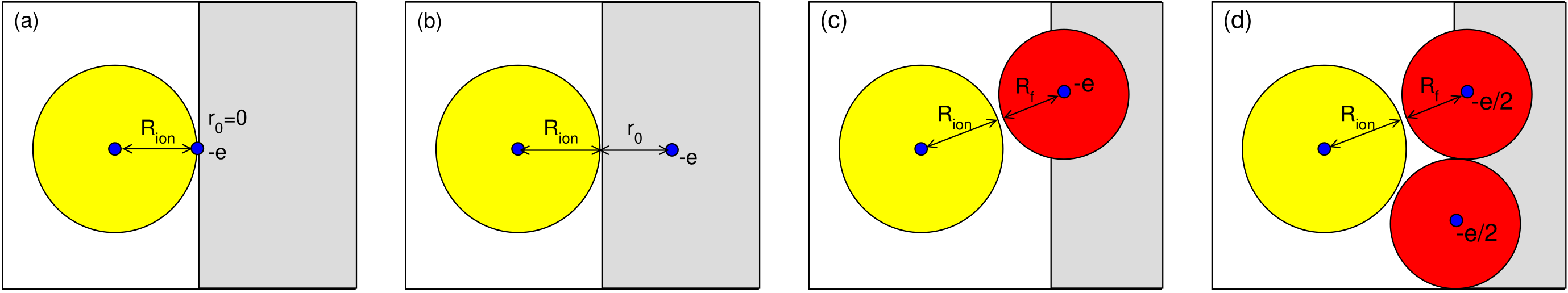}
	\caption{\small The four models for localizing the charges of the COO$^{-}$ groups (a) the $-e$ point charge is fixed on the surface of the pore ($r_{0}=0$), the model used in our previous work~\cite{fabian_jml_2022}; (b) the $-e$ point charge is behind the surface at $r_{0}$ distance from the surface; (c) the $-e$ point charge is located in the center of a hard sphere of radius $R_{\mathrm{f}}$ (this model is more appropriate for silanol groups of silica nanopores); (d) the $-e$ charge is distributed between two oxygen atoms in the form of $-e/2$ partial charges. For the latter two models, the oxygen atoms are confined to positions at distance $r_{0}$ behind the surface using the harmonic potential of Eq.~\ref{eq:harmonic}. For more details, see Fig.~\ref{fig3}. }
	\label{fig2}
\end{figure*}

The pore wall carries a negative surface charge density, $\sigma = -1\ e/\mathrm{nm}^2$, which is consistent with experimental observations \cite{gillespie_bj_2008_nanopore}.
In our previous work~\cite{fabian_jml_2022}, we distributed fractional point charges $Q$ on a lattice on the inner pore wall such that the prescribed surface charge density, $\sigma$, was maintained.
Specifically, the grid spacing along the $z$ axis was $\Delta x$ (so that $H/\Delta x$ is an integer, $H=12$ nm), while the spacing in the azimuthal direction was $\Delta\phi$ (so that $2\pi/\Delta\phi$ is an integer and $\Delta x \approx R\Delta\phi$, $R=2.7$ nm). 
The magnitude of each surface point charge was then given by $Q = \sigma\,\Delta x R \Delta \phi $. 
The surface charges were immobile and fixed at the pore wall.

\begin{figure}[t]  
	\centering
	\includegraphics[width=0.8\linewidth]{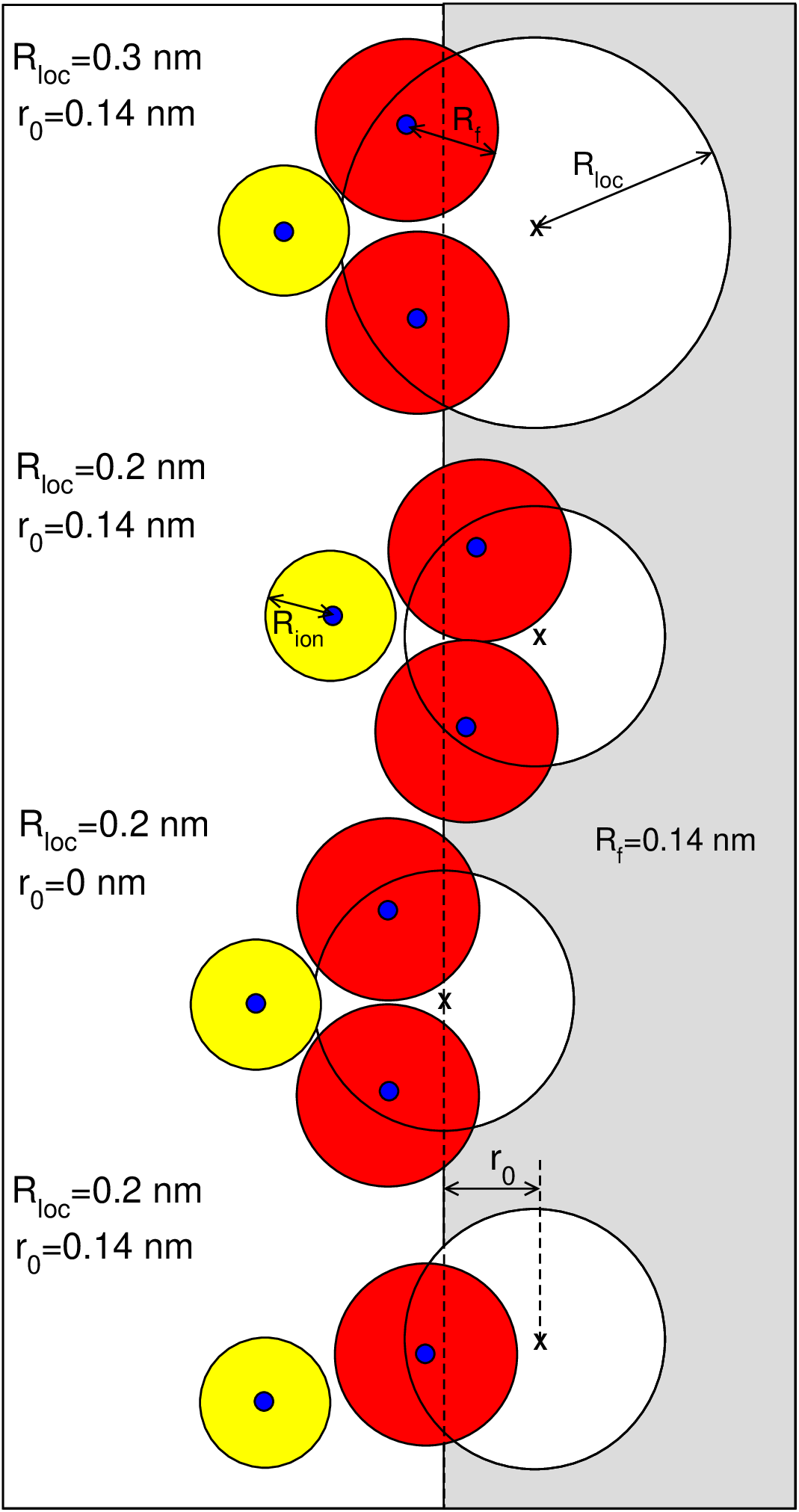} 
	\caption{\small Variations of modeling the oxygen atoms considered in this work. The red oxygen atoms represented as charged hard spheres of radius $R_{\mathrm{f}}$ are localized by a harmonic potential (Eq.~\ref{eq:harmonic}) with a reference position that is at distance $r_{0}$ from the wall (behind it). The radius of the confinement is $R_{\mathrm{loc}}$. From top to bottom: 
		(i) the $R_{\mathrm{loc}}=0.3$ nm value allows larger mobility for the oxygen atoms hanging at the end of a relatively long side chain; 
		(ii) the value $R_{\mathrm{loc}}=0.2$ nm allows smaller mobility for the oxygen atoms that are bound to the pore material more tightly; 
		(iii) the value $r_{0}=0$ allows endgroups to protrude more deeply into the electrolyte; 
		(iv) the group charge is located on a single oxygen atom.}
	\label{fig3}
\end{figure}

While the effect of the lattice spacing, $\Delta x$, was investigated in our previous work~\cite{fabian_jml_2022}, we revisit this issue here by examining the impact of varying $\Delta x$ within the fitting procedure described below.
Then we fix $\Delta x = 1$ nm (and $Q=-e$), a value that corresponds to the experimental situation in which the wall charges are associated with COO$^{-}$ groups on the surface of a PET nanopore~\cite{gillespie_bj_2008_nanopore,he_jacs_2009}.
Figure \ref{fig2} shows three possibilities for modeling the functional groups that differ in complexity.

The model in Fig.~\ref{fig2}c represents the atom carrying the surface point charge explicitly by a charged hard sphere (just like the ions). 
The radius of the atom is $R_{\mathrm{f}}=0.14$ nm, that of an oxygen atom.
Then the DCA of an ion of radius $R_{\mathrm{ion}}$ to the surface charge is the sum of the two radii: $\mathrm{DCA}=R_{\mathrm{ion}}+R_{\mathrm{f}}$.
In a more detailed version of this model the charge is split between two oxygen atoms (Figure \ref{fig2}d).

We model these oxygens following the description of Finnerty et al.~\cite{Finnerty}. 
Accordingly, the oxygen atom (located at position $\mathbf{r}$) is confined to a region around a reference position $\mathbf{r}_{0}$ by a harmonic potential:
\begin{equation}
	u_{\mathrm{loc}}(\mathbf{r}) =
	\begin{cases}
		\infty, & \text{for } |\mathbf{r}-\mathbf{r}_{0}| > R_{\mathrm{loc}}, \\[6pt]
		-\dfrac{k_{\mathrm{loc}}}{R^2_{\mathrm{loc}}}\,\bigl|\mathbf{r}-\mathbf{r}_{0}\bigr|^2, & \text{for } |\mathbf{r}-\mathbf{r}_{0}| \le R_{\mathrm{loc}},
	\end{cases}
	\label{eq:harmonic}
\end{equation}
with effective spring constant $k_{\mathrm{loc}}/R_{\mathrm{loc}}^2$.
The value $k_{\mathrm{loc}}=4.5 kT$ was used in this work; using other value for $k_{\mathrm{loc}}$ does not influence the final result. 
Here, $R_{\mathrm{loc}}$ is the maximum allowable distance of the atom from $\mathbf{r}_{0}$; the oxygen atoms' centers are constrained by the harmonic potential to move only within a sphere of this radius and cannot overlap with each other.
Finnerty et al.~\cite{Finnerty} developed this model as a reduced representation of amino-acid side chains in ion channels, based on the scheme proposed by Yu et al. \cite{Yu-et-al}. 

Thus, the surface groups modeled this way are not rigid; some movement of oxygen atoms mimicking vibration is allowed.
Various models regarding the number of oxygen atoms, their locations ($r_0$), and the value of $R_{\mathrm{loc}}$ are shown in Fig.~\ref{fig3} and will be investigated in this study.

We also use a ``behind the wall'' model (Fig.~\ref{fig2}b), where the surface point charge $-e$ is placed behind the hard surface at distance $r_0$.
A special case of this model is when we place the charges on the surface ($r_{0}=0$, Fig.~\ref{fig2}a).

\subsection{NP+LEMC method}

The driving force for drift and diffusion is the gradient of the electrochemical potential in the Nernst-Planck (NP) equation:
\begin{equation}
 \mathbf{j}_{i}(\mathbf{r})= -\frac{1}{kT} D_{i}(\mathbf{r})c_{i}(\mathbf{r})\nabla \mu_{i}(\mathbf{r}) .
\label{eq:NP}
\end{equation}
In this equation, $i$ is the ion species,  $k$ is the Boltzmann constant, $T$ is the temperature ($T=298. 15$ K), $\mathbf{j}_{i}(\mathbf{r})$ is the particle current density, $D_{i}(\mathbf{r})$ is the diffusion coefficient profile, $c_{i}(\mathbf{r})$ is the concentration profile, and $\mu_{i}(\mathbf{r})$ is the electrochemical potential profile.

The diffusion constant profile is constructed as a step function, where its value for a given ion species is a constant outside the pore ($D_{i}^{\mathrm{b}}$, see Table \ref{tab}) and a different constant ($D_{i}^{\mathrm{p}}$) inside the pore:
\begin{equation}
D_i (z) = \left\lbrace 
\begin{array}{ll}
D_i^{\mathrm{p}} & \quad \mathrm{for} \quad |z|< H/2 \\
 D_i^{\mathrm{b}}  & \quad  \mathrm{for} \quad |z|\geq H/2\\
\end{array}
\right. 
\end{equation}
where $H=12$ nm is the length of the pore.
The value $D_{i}^{\mathrm{p}}$ is used as a fitting parameter (i.e., its value is fitted so that the calculated conductivity matches the experimental value).

To solve the NP equation, a closure is needed between the concentration profile, $c_i (\mathbf{r})$, and the chemical potential profile, $\mu_i (\mathbf{r})$.
In Poisson-Nernst-Planck theory, this closure is provided by Poisson-Boltzmann (PB) theory.
Since the PB theory does not describe ionic correlations beyond the mean-field approximation (which are important in this study)	, we use a computer simulation approach, the Local Equilibrium Monte Carlo (LEMC) method~\cite{boda_jctc_2012} to obtain the closure.

The LEMC method is the non-equilibrium variant of the Grand Canonical Monte Carlo method in which the chemical potential is not constant globally, but is constant in small volume elements in which local thermodynamic equilibrium is assumed.
The LEMC simulation provides the $c_i (\mathbf{r})$ profile as an output to the input $\mu_i (\mathbf{r})$ profile.
The NP equation must be solved together with the LEMC simulation in a self-consistent way (NP+LEMC) such that the continuity equation describing the conservation of mass, $\nabla \cdot \mathbf{j}_{i}(\mathbf{r})=0$, is also satisfied.
An iteration method is used to couple the LEMC simulation to the NP transport equation (NP+LEMC method)~\cite{boda_jctc_2012}.

In the LEMC simulations we employ 3D geometry, with discrete surface charges as described. 
However, to make the coupling with the NP equation computationally tractable, we assume rotational symmetry. 
That is, after the 3D LEMC simulation, those results are averaged over $\phi$ and put into the NP equation whose output is rotationally symmetric chemical potentials, $\mu_{i}(z,r)$, used in the next LEMC simulation.
Then, we divide the $z,r$ plane into $\Delta z \times \Delta r$ rectangles, so that the volume elements are rings with this rectangular cross section ($\Delta z, \Delta r \approx 0.2$ nm).

The total flux through the cross-section $A$ of the pore is obtained by integrating the current density obtained from Eq.~\ref{eq:NP} over the cross section: 
\begin{equation}
 J_{i}=\int_{A}\mathbf{j}_{i}\cdot \mathrm{d}\mathbf{a} .
\label{eq:flux}
\end{equation}
The total electrical current flowing through the pore is $I=\sum_{i}q_{i}J_{i}=\sum_{i}I_{i}$, where $I_{i}$ is the electrical current carried by ionic species $i$. 
The conductivity of the pore is $G=I/U$, while the conductivity for a given ion is $G_i =I_i/U$ (the $I$ vs.\ $U$ curves are closely linear). 

\subsection{Fitting the diffusion coefficients in the pore}

The experimental data that is available for us is the conductance, $G$, of the pore as a function of the mole fraction of CaCl$_{2}$, 
\begin{equation}
	\eta = \dfrac{[\mathrm{CaCl}_{2}]}{\mathrm{[KCl]+[CaCl}_{2}]}
\end{equation}
in a KCl+CaCl$2$ mixture~\cite{gillespie_bj_2008_nanopore}. 
Results for NaCl and LiCl are also available.
  
In our previous paper~\cite{fabian_jml_2022}, our strategy was to fit $D_{\mathrm{K}^{+}}^{\mathrm{p}}$ and  $D_{\mathrm{Ca}^{2+}}^{\mathrm{p}}$ to the endpoints $\eta=0$ and $1$, respectively.
Our goal was to reproduce the $G(\eta)$ curve in between, an effort that was not entirely successful.
That approach, however, leaves a third unknown parameter, the pore diffusion coefficient of Cl$^{-}$, $D_{\mathrm{Cl}^{-}}^{\mathrm{p}}$.

We solved this problem by assigning a $D_{\mathrm{Cl}^{-}}^{\mathrm{p}}$ value to the chloride ions by tying that value to either that of K$^{+}$ or Ca$^{2+}$ so that 
\begin{equation}
	\dfrac{D_{\mathrm{Cl}^{-}}^{\mathrm{p}}}{D_{\mathrm{Cl}^{-}}^{\mathrm{b}}} = 	
	\dfrac{D_{\mathrm{K}^{+}}^{\mathrm{p}}}{D_{\mathrm{K}^{+}}^{\mathrm{b}}} \quad  \mathrm{or} \quad
	\dfrac{D_{\mathrm{Cl}^{-}}^{\mathrm{p}}}{D_{\mathrm{Cl}^{-}}^{\mathrm{b}}} = 
	\dfrac{D_{\mathrm{Ca}^{2+}}^{\mathrm{p}}}{D_{\mathrm{Ca}^{2+}}^{\mathrm{b}}};	
	\label{eq:tying}
\end{equation}
that is, the ratio of diffusion constants in the pore and in the bulk agree.
The choice (fitting to K$^{+}$ or Ca$^{2+}$) did not influence the total current, it just influenced the ratio of Ca$^{2+}$ and Cl$^{-}$ currents~\cite{fabian_jml_2022}.
In both cases, we obtained smaller conductance for Cl$^{-}$ than for Ca$^{2+}$ for $\eta=1$ (pure CaCl$_{2}$), a result that was consistent with a negatively charged pore that was expected to be cation selective on the basis of classical double layer theories like PB.
	 
Our new molecular dynamics (MD) simulations~\cite{shabbir_jcp_2026} for CaCl$_{2}$ transport through a negatively charged silica nanopore, however, provided results that implied a non-selective, or rather Cl$^{-}$ selective pore (depending on the force field used). 
In that study, the oxygens of the silanol groups of the silica nanopore attracted the Ca$^{2+}$ ions so strongly that the Ca$^{2+}$ ions were practically immobile (they had a small mobility in the surface layer near the pore wall), while the Cl$^{-}$ ions were not present in the surface layer.
Consequently, the surface layer did not contribute to the conduction, only the volume region along the centerline.
This region, however, shows bulk-like behavior, where Cl$^{-}$ current is larger than Ca$^{2+}$ current because its diffusion constant is larger (see Table \ref{tab}).

Furthermore, this new result qualitatively agrees with the experimental observations of He et al.~\cite{he_jacs_2009} for rectifying conical PET nanopores, where measurements with only KCl showed cation selectivity, CoSepCl$_{3}$ (a 3:1 electrolyte) showed anion selectivity, and CaCl$_{2}$ displayed intermediate, nearly non-selective behavior.

Motivated by these facts, here we use a modified approach of fitting to three points: pure KCl ($\eta=0$), pure CaCl$_{2}$ ($\eta=1$), and the minimum of the experimental $G(\eta)$ curve at $\eta=0.2$.
In the fitting procedure, we use the result that in the LEMC simulations the conductance depends linearly on $D_{i}^{\mathrm{p}}$ because these diffusion coefficient values are very small (data not shown).
Therefore, we define the ratio $G_{i}'(\eta) = G_{i}(\eta)/D_{i}^{\mathrm{p}}$ that is an archetypal normalized conductance versus $\eta$ curve that is independent of the fitted diffusion coefficients, a fact we use frequently below.
The $G_{i}'(\eta) $ values can be obtained from a single NP+LEMC simulation due to the linearity.
 
The fitted diffusion coefficients in the pore, $D_{i}^{\mathrm{p}}$, are obtained from the following equations
\begin{widetext}
\begin{alignat}{7}
	G^{\mathrm{exp}}(0)  \:\: =&  & & \:\: D_{\mathrm{Cl}^{-} }^{\mathrm{p}}G'_{\mathrm{Cl}^{-} }(0) 
	                           &+& \:\: D_{\mathrm{K} ^{+} }^{\mathrm{p}}G'_{\mathrm{K} ^{+} }(0)  \nonumber \\
	G^{\mathrm{exp}}(0.2)\:\: =&  \:\: D_{\mathrm{Ca}^{2+}}^{\mathrm{p}}G'_{\mathrm{Ca}^{2+}}(0.2) \:\:  
						       &+& \:\: D_{\mathrm{Cl}^{-} }^{\mathrm{p}}G'_{\mathrm{Cl}^{-} }(0.2) \:\:
						       &+& \:\: D_{\mathrm{K} ^{+} }^{\mathrm{p}}G'_{\mathrm{K} ^{+} }(0.2)  \nonumber \\
    G^{\mathrm{exp}}(1) \:\:  =&  \:\:D_{\mathrm{Ca}^{2+}}^{\mathrm{p}}G'_{\mathrm{Ca}^{2+}}(1)  
    					       &+& \:\:D_{\mathrm{Cl}^{-} }^{\mathrm{p}}G'_{\mathrm{Cl}^{-} }(1) 
    					       & &
    \label{eq:fit-eq}
\end{alignat}
\end{widetext}
where $G^{\mathrm{exp}}(\eta)$ is the experimental value for the total conductance. 

Although now we are fitting to three points, as we will see, there is space for the $G(\eta)$ curves to behave quite differently between the fitting points.
The questions that we now seek the answers to are:
\begin{enumerate}
	\item How do the $G(\eta)$ curves behave in between the three fitting points ($0<\eta<0.2$ and $0.2<\eta<1$)?
	\item How do the simulations agree with experimental and MD findings that the Cl$^{-}$ conductance is larger than Ca$^{2+}$ conductance ($G_{\mathrm{Cl}^{-}}(1)>G_{\mathrm{Ca}^{2+}}(1)$) for pure CaCl$_{2}$ (bulk-like behavior)?
	\item How do the models of the carboxyl groups change the $G(\eta)$ curves?
\end{enumerate}

\section{Results and Discussion}
\label{sec:results}

We begin with the simplest model in which fixed point charges $Q$ are placed on a $\Delta x\ \times \Delta x$ grid behind the pore wall at a distance $r_0$ from the surface (Fig.~\ref{fig1}b).

\subsection{Changing localization}

First, we revisit the effect of varying $\Delta x$ (and, simultaneously, $Q$, to keep the average surface charge constant) using the three-point fitting procedure defined in Eq.~\ref{eq:fit-eq}, with the surface charges located at the pore wall ($r_0 = 0$). 
Our previous work showed that the minimum in the total conductance (obtained from Eq.~\ref{eq:tying}) deepens as $\Delta x$ increases; however, the overall shape of the resulting curve did not quantitatively match the experimental data (see Fig.~11 of Ref.~\cite{fabian_jml_2022}). 
Since the parameter $\Delta x$ controls the strength of charge inversion through the degree of charge localization, it is natural to ask how the application of the three-point fitting procedure (Eq.~\ref{eq:fit-eq}) affects the resulting conductance curves in the present analysis.

\begin{figure}[t!]  
	\centering
	\includegraphics[width=0.99\linewidth]{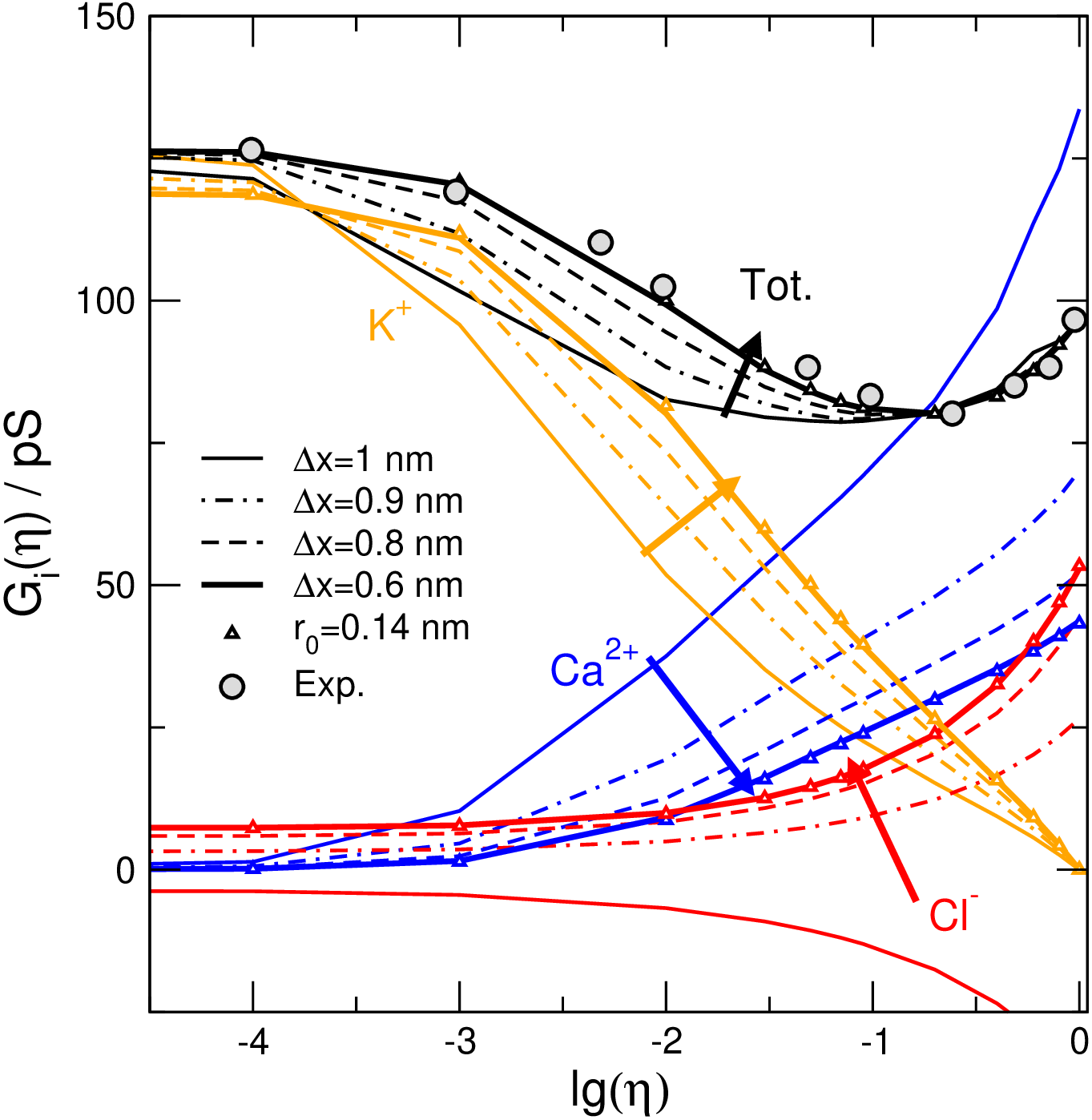} 
	\caption{\small Conductances of ionic species (Ca$^{2+}$, K$^{+}$, and Cl$^{-}$) and total conductance as functions of Ca$^{2+}$ mole fraction, $\eta$, for different models of carboxyl groups. Lines refer to fixed-charge models with the point charge being ``on the wall'' ($r_0=0$) with $ \Delta x$ grid spacing and reduced point charges $Q$ to maintain $\sigma=-1$ $e$/nm$^{2} $. The arrows show the direction of increasing $\Delta x$. The small open triangles show the results obtained for modeling the carboxyl groups point charges at $r_0 =0.14$ nm distance ``behind the wall'' using $\Delta x=1$ nm spacing and $Q=-e$. The large gray circles denote experimental results~\cite{gillespie_bj_2008_nanopore}. Total concentration is fixed at $0.1$ M.}
	\label{fig4}
\end{figure}

Figure \ref{fig4} shows he ionic conductances, $G_{i}$, as functions of $\lg(\eta)$ for different values of $\Delta x$ with various line styles ($r_0 =0$) as obtained from Eq.~\ref{eq:fit-eq}. 
Arrows indicate increasing $\Delta x$.
The total conductance increases with increasing $\Delta x$ (black curves) in the $0<\eta <0.2$ region, and becomes increasingly similar to the experimental data (large gray spheres).

For localization $\Delta x=0.6$ nm, the curve (thick line in Fig.~\ref{fig4}) practically becomes identical with the curve for $\Delta x=1$ nm, $Q=-e$, $r_0 = 1.4$ nm (small open triangles). 
This latter model (discussed in the next subsection) decreases charge inversion in a different way: placing the $-e$ point charges ``behind the wall'' at distance $r_0$, as compared to reducing $Q$.
This result implies that the strength of charge inversion can be manipulated in two distinct ways: (1) changing localization (characterized by the $\Delta x$ control parameter) and (2) changing the DCA (characterized by the $r_0$ control parameter).

The $G_i (\eta)$ curves for the individual ions show similar behavior as functions of the control parameters: as functions of $\Delta x$ in Fig.~\ref{fig4} and as functions of $r_0$ in Fig.~\ref{fig5}.
Therefore, we refer the detailed analyses of this behavior to the next subsection where we discuss that model.
Here, we just note that charge inversion is so strong in the $\Delta x=1$ nm, $r_0 =0$ nm case that fitting on the basis of Eq.~\ref{eq:fit-eq} provides unphysical results: too large Ca$^{2+}$ conductance and negative Cl$^{-}$ conductance.

\subsection{Changing distance of closest approach}

We continue with the model in which fixed $-e$ point charges are placed on a $1\; \mathrm{nm} \times 1 \; \mathrm{nm}$ grid ($\Delta x = 1$ nm, $Q=-e$) behind the pore wall at a distance $r_0$ from the surface (Fig.~\ref{fig1}b). 
The case $r_0 =0$ (the charge is on the wall, Fig.~\ref{fig1}a) is the point where our previous work~\cite{fabian_jml_2022} and the previous subsection concluded.
The value $r_0=R_{\mathrm{f}}=0.14$ nm (triangles in Fig.~\ref{fig4}), on the other hand, represents the case where the DCA between the ionic charge centers and the surface charges coincides with that obtained when carboxyl groups are modeled explicitly by oxygen atoms (Figs.~\ref{fig1}c-d and Fig.~\ref{fig2}).
In addition to these two limiting cases, we investigate intermediate cases with $0\le r_0\le R_{\mathrm{f}}$.

\begin{figure}[t!]  
	\centering
	\includegraphics[width=0.99\linewidth]{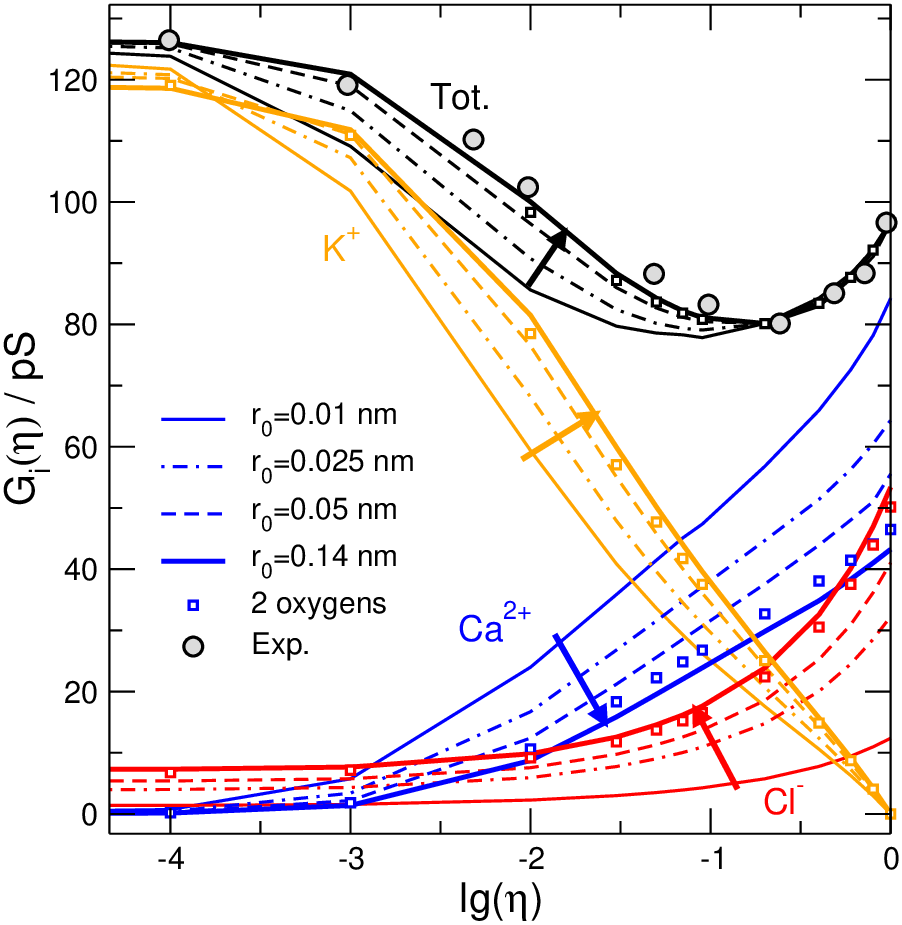}
	\caption{\small Conductances of ionic species (Ca$^{2+}$, K$^{+}$, and Cl$^{-}$) and total conductance as functions of Ca$^{2+}$ mole fraction, $\eta$, for different models of carboxyl groups. Lines refer to fixed-charge models with the point charge being ``behind the wall'' at distance $r_0$ from the wall. The arrows show the direction of increasing $r_0$. The small open squares show the results obtained for modeling the carboxyl groups with two oxygen atoms (charged hard spheres with radius $R_\mathrm{f}=0.14$ nm) each carrying $-0.5e$ charge ($r_0 =0.14$ nm, $R_{\mathrm{loc}}=0.2$ nm). The large gray circles denote experimental results~\cite{gillespie_bj_2008_nanopore}. Total concentration is fixed at $0.1$ M.}
	\label{fig5}
\end{figure}

Figure~\ref{fig5} shows the ionic conductances, $G_{i}$, (as obtained from Eq.~\ref{eq:fit-eq}) as functions of $\lg(\eta)$ for different values of $r_0$, with arrows indicating increasing $r_0$. 
As $r_0$ increases, the Ca$^{2+}$ conductance decreases, while the Cl$^{-}$ and K$^{+}$ conductances increase. 
The total conductance, $G=\sum_{i}G_{i}$, for which experimental data are available, likewise increases in the $0<\eta<0.2$ region and approaches the experimental curve as $r_0\rightarrow R_{\mathrm{f}}=0.14$ nm.

\begin{figure}[t]  
	\centering
	\includegraphics[width=0.93\linewidth]{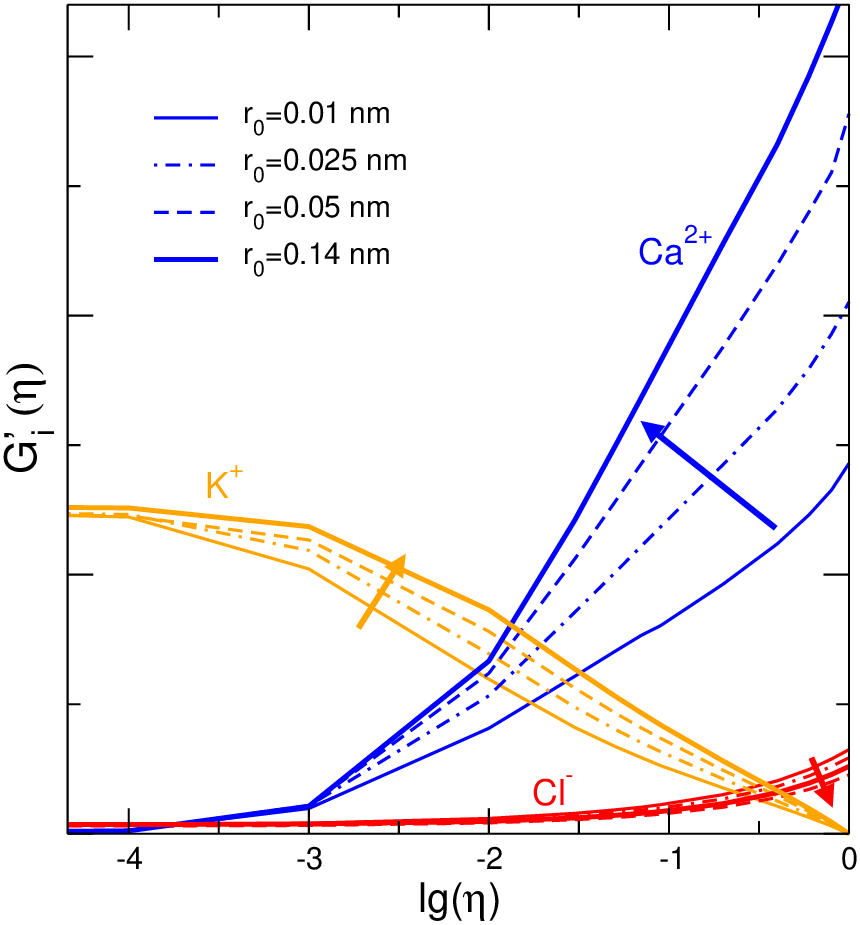}
	\caption{\small The $G_{i}'$ normalized conductances for ionic species Ca$^{2+}$, K$^{+}$, and Cl$^{-}$ as functions of Ca$^{2+}$ mole fraction, $\eta$, for different models of carboxyl groups. Lines refer to fixed-charge models with the point charge being ``behind the wall'' at distance $r_0$ from the wall. The arrows show the direction of increasing $r_0$. The $G_{i}'$ values are obtained by dividing the conductances ($G_{i}$) of Fig.~\ref{fig5} with the diffusion constants ($D_{i}^{\mathrm{p}}$) of Fig.~\ref{fig7}A. The values of $G_{i}$ are not shown; we focus on the trends.}
	\label{fig6}
\end{figure}

\begin{figure}[t]  
	\centering
	\includegraphics[width=0.9\linewidth]{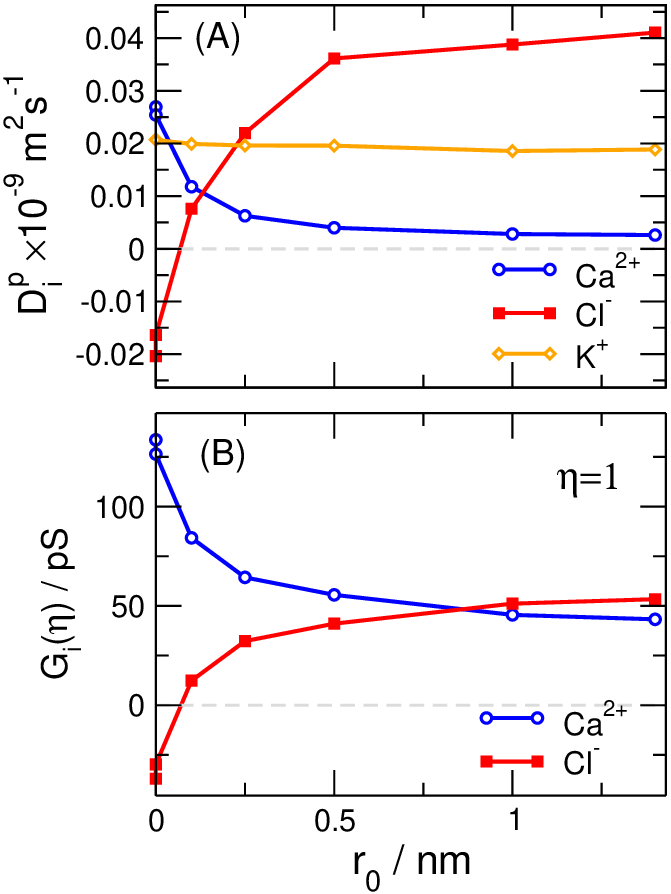}
	\caption{\small (A) Diffusion constants of ionic species Ca$^{2+}$, K$^{+}$, and Cl$^{-}$ (from Eq.~\ref{eq:fit-eq}) as functions of $r_0$ in the framework of the fixed-charge model with the point charge being ``behind the wall'' at distance $r_0$ from the wall. (B) Ca$^{2+}$ and Cl$^{-}$ conductances as functions of $r_0$ for pure CaCl$_{2}$ ($\eta =1$). }
	\label{fig7}
\end{figure}

\begin{figure*}[t!]  
	\centering
	\includegraphics[width=\linewidth]{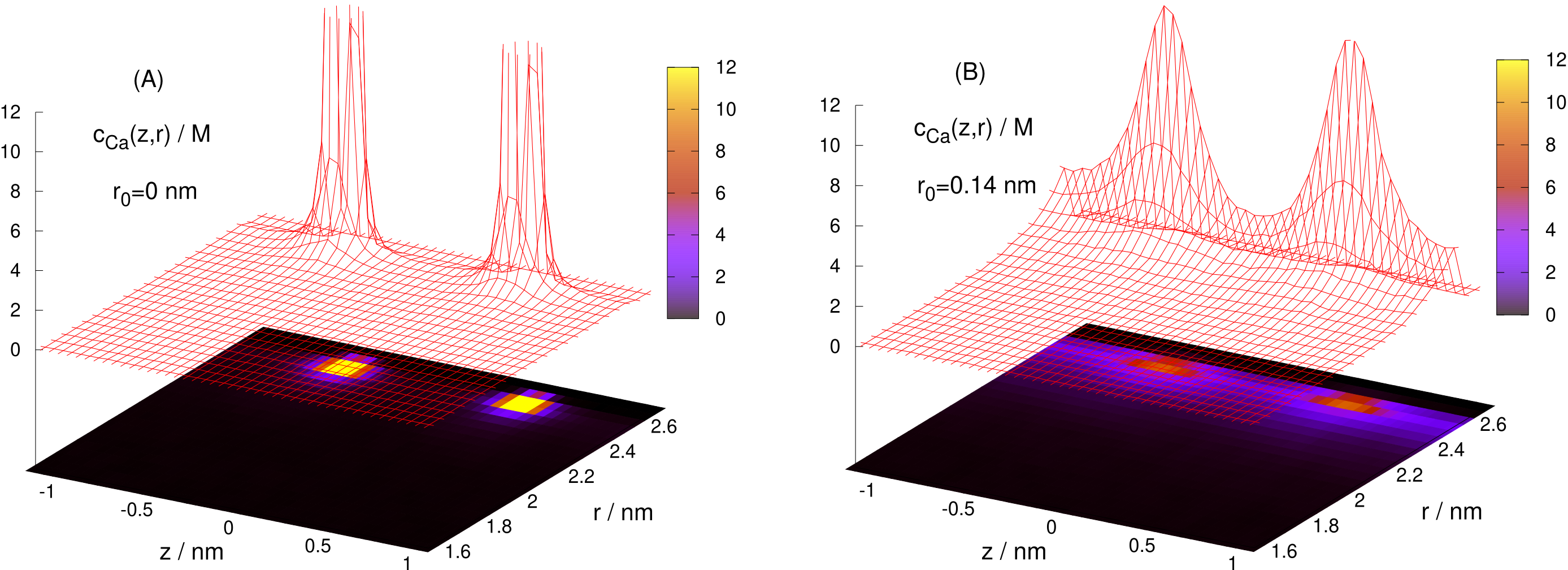}
	\caption{\small Concentration profile of ionic species Ca$^{2+}$ as a function $z$ and $r$ in the framework of the fixed-charged model for (A) $r_0 =0$ nm and (B) $r_0 =0.14$ nm at CaCl$_{2}$ mole fraction $\eta = 0.2$. The value of the large peaks in panel (A) is about $104$ M. }
	\label{fig8}
\end{figure*}

\begin{figure}[t]  
	\centering
	\includegraphics[width=0.8\linewidth]{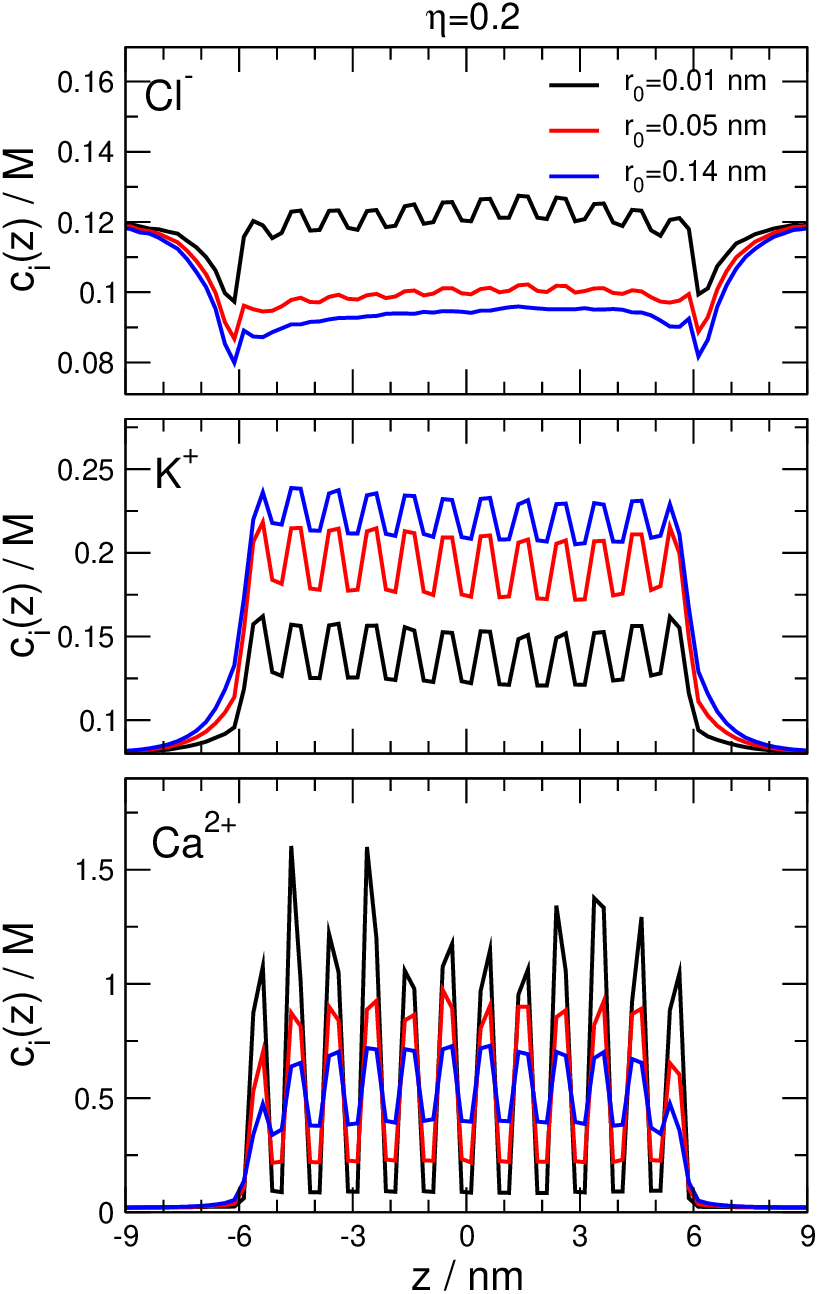}
	\caption{\small Axial concentration profiles of ionic species Ca$^{2+}$, K$^{+}$, and Cl$^{-}$ for different values of $r_0$ in the framework of the fixed-charge model at CaCl$_{2}$ mole fraction $\eta = 0.2$.}
	\label{fig9}
\end{figure}

This behavior arises from the simultaneous change of both $G_{i}'$ and $D_{i}^{\mathrm{p}}$ (Figs.~\ref{fig6} and \ref{fig7}A).
$G_{i}'$ is determined by the LEMC simulation only, so it depends on the state point (determined by $\eta$) and on the local interactions (tuned by $r_0$) as shown in Fig.~\ref{fig6}. 
The conductances, on the other hand, change because the changing $G_{i}'(\eta)$ coefficients in Eq.~\ref{eq:fit-eq} lead to different $D_{i}^{\mathrm{p}}$ solutions (Fig.~\ref{fig7}A), and, therefore, different $G_{i}(\eta)$ values between the fitting points (Fig.~\ref{fig5} and \ref{fig7}B).
Varying $r_0$ tunes the strength of charge inversion, so it changes the concentration profiles (Figs.~\ref{fig8} and \ref{fig9}), and, consequently, the $G_i'$ variables (Fig.~\ref{fig6}).

The behavior of Ca$^{2+}$ is strongly affected by charge inversion (see the peaks and depletion zones in Figs.~\ref{fig8} and bottom panel of \ref{fig9}) so the $G_{\mathrm{Ca}}'(\eta)$ function depends sensitively on $r_0$: Ca$^{2+}$ ions become more mobilized with increasing $r_0$ ($G_{\mathrm{Ca}}'$ increases, see blue curves in Fig.~\ref{fig6}).
The conductance of Ca$^{2+}$, on the other hand, decreases with increasing $r_0$ (blue curves in Fig.~\ref{fig5}).

The $G_{\mathrm{Cl}}'(\eta)$ function, on the other hand, is small and decreases with increasing $r_0$ (red curves in Fig.~\ref{fig6}).
$G_{\mathrm{Cl}}'$ is small because Cl$^{-}$ concentrations are smaller in the pore compared to cations (see the Cl$^{-}$ concentration profiles in the top panel of Fig.~\ref{fig9}); it is a negatively charged pore after all.	
$G_{\mathrm{Cl}}'$ decreases due to decreasing Cl$^{-}$ concentration with increasing $r_0$ due to lessened charge inversion (less overcharged pore wall draws less co-ions).
The conductance of Cl$^{-}$ increases with increasing $r_0$ (red curves in Fig.~\ref{fig5}) despite this concentration decrease because a Cl$^{-}$ conductance is required to reproduce the data.

The $G_{\mathrm{K}}'(\eta)$ function shows an increasing trend with increasing $r_0$ (orange curves in Fig.~\ref{fig6}) because the quantity of K$^{+}$ increases in the pore as charge inversion weakens (see the K$^{+}$ concentration profiles in the middle panel of Fig.~\ref{fig9}).
The conductance profiles show the same trend (orange curves in Fig.~\ref{fig5}).

The above discussion can readily be applied to the $\Delta x$-dependence in the previous subsection by replacing ``increasing $r_0$'' with ``decreasing $\Delta x$'' since both weaken charge inversion.
Axial concentration profiles for different values of $\Delta x$ are found in Fig.~10 of Ref.~\cite{fabian_jml_2022}.
Detailed discussion about the role of localization is found in Ref.~\cite{boda_entropy_2020}.

This complex behavior is the outcome of (1) simulations to get $G_{i}'$ and (2) fitting on the basis of Eq.~\ref{eq:fit-eq} to get $D_{i}^{\mathrm{p}}$ and $G_{i}$.
The question now is which model of the surface charge (which $r_0$) produces results that are in agreement with the experimental $G_{i}(\eta)$ profiles between the fitting points and which model is in agreement with other considerations such as experiment for pure CaCl$_{2}$~\cite{he_jacs_2009}, MD simulations for pure CaCl$_{2}$~\cite{shabbir_jcp_2026}, and physical intuition. 

The opposite $r_0$-dependence of $G_{i}$ and $G_{i}'$ for Ca$^{2+}$ and Cl$^{-}$ is the result of the $D_{\mathrm{Ca}}^{\mathrm{p}}$ and $D_{\mathrm{Cl}}^{\mathrm{p}}$ values showing opposite trends as functions of $r_0$ (Fig.~\ref{fig7}A).
The case of $r_0 =0$ (charge on the wall) is physically not meaningful first of all because $D_{\mathrm{Cl}}^{\mathrm{p}}$ is negative, and because $D_{\mathrm{Ca}}^{\mathrm{p}}$ is larger than $D_{\mathrm{K}}^{\mathrm{p}}$, which is unlikely given that $D_{\mathrm{Ca}}^{\mathrm{b}}<D_{\mathrm{K}}^{\mathrm{b}}$ in the bulk (Table \ref{tab}) and that negative surface charges further immobilize Ca$^{2+}$ ions.
Therefore, we do not obtain physically sensible results if charge inversion is too strong.

As $r_0$ is increased (and charge inversion decreased), the $D_{\mathrm{Ca}}^{\mathrm{p}}$ curve declines steeply below $D_{\mathrm{K}}^{\mathrm{p}}$, while $D_{\mathrm{Cl}}^{\mathrm{p}}$ increases steeply, first becoming positive and then even surpassing the $D_{\mathrm{K}}^{\mathrm{p}}$ value. 
As we increase $r_0$ further, the the $D_{\mathrm{Ca}}^{\mathrm{p}}$ and the $D_{\mathrm{Cl}}^{\mathrm{p}}$ values stabilize.
The $D_{\mathrm{Cl}}^{\mathrm{p}}$ value in the $r_0 \to 0.14$ nm limit is about twice as large as $D_{\mathrm{K}}^{\mathrm{p}}$, which implies that Cl$^{-}$ ions have larger mobility than K$^{+}$ ions.
This makes sense, because the negative pore charges attract K$^{+}$ ions limiting their mobility, while Cl$^{-}$ ions experience the opposite effect.
The $D_{\mathrm{Ca}}^{\mathrm{p}}$ value is about $7$ times smaller than $D_{\mathrm{K}}^{\mathrm{p}}$, which implies that Ca$^{2+}$ ions have smaller mobility than K$^{+}$ ions. 
This also makes sense, because Ca$^{2+}$ ions correlate more strongly with the surface groups than K$^{+}$ ions.
What is more, they overcharge the COO$^{-}$ group forming a positive complex that repels K$^{+}$.

The relative magnitudes of Ca$^{2+}$ and Cl$^{-}$ conductances in pure CaCl$_2$ are of particular interest, as both experimental~\cite{he_jacs_2009} and simulation~\cite{shabbir_jcp_2026} studies indicate that negatively charged nanopores exhibit bulk-like behavior in this limit. 
Specifically, the ratio $G_{\mathrm{Ca}^{2+}}/G_{\mathrm{Cl}^{-}}$ is expected to be similar to its bulk value, $2D_{\mathrm{Ca}^{2+}}^{\mathrm{b}}/D_{\mathrm{Cl}^{-}}^{\mathrm{b}}\approx0.78$ (the factor $2$ is due to the double charge of Ca$^{2+}$), implying that Cl$^{-}$ conductance may exceed that of Ca$^{2+}$. 
Figures~\ref{fig5} and~\ref{fig7}B show that this behavior is qualitatively reproduced when $r_0=R_{\mathrm{f}}$.

As discussed above, the $G_{i}'$ values essentially depend on the concentration profiles.
Ca$^{2+}$ ions are strongly adsorbed to the pore charges, producing pronounced concentration peaks and depletion zones between them (Fig.~\ref{fig8}). 
Because the pore charges are arranged in rings separated by $\Delta x$ spacing, sharp concentration peaks for Ca$^{2+}$ appear at the ring positions, separated by depletion zones.
The peaks are much higher and depletion zones are deeper for $r_0 = 0$ nm (Fig.~\ref{fig8}A) than for $r_0 = 0.14$ nm (Fig.~\ref{fig8}B). 
Averaging these concentration profiles over $r$, we obtain the axial concentration profiles shown in Fig.~\ref{fig9}.
When the pore is viewed as a series of resistive elements along the axial direction, the low-concentration depletion zones act as high-resistance segments that make the resistance of the whole circuit high.
The physical mechanism behind the depletion zone having high resistance is that Ca$^{2+}$ movement between localized binding sites (e.g., across depletion zones) is a rare event that limits Ca$^{2+}$ current.
This is what makes the $G_{i}’$ the primary driver of conductance, as it reflects the axial concentration profiles.

For small $r_0$, fitting to the three experimental data points yields unrealistically large values of $D_{\mathrm{Ca}^{2+}}^{\mathrm{p}}$, accompanied by unrealistically small---and even negative---values of $D_{\mathrm{Cl}^{-}}^{\mathrm{p}}$ near $r_0\approx 0$. 
To disentangle the roles of mobility and concentration effects, we plot the $G_i'(\eta)$ curves in Fig.~\ref{fig6}.
For $r_0=0.01$ nm, corresponding to strong charge inversion, $G_{\mathrm{Ca}^{2+}}'$ is small due to the deep depletion zones (black line in bottom panel of Fig.~\ref{fig9}).
On the other hand, $G_{\mathrm{K}^{+}}'$ is also small because the average concentration K$^{+}$ ions is small (middle panel of Fig.~\ref{fig9}).	

As $r_0$ increases, charge inversion weakens; Ca$^{2+}$ concentration peaks decrease in magnitude and the depletion zones become shallower (see Fig.~\ref{fig8}B and the blue curve in the bottom panel of Fig.~\ref{fig9} for $r_0 =0.14$ nm). 
As a result, $G_{\mathrm{Ca}^{2+}}'$ increases with increasing $r_0$ (Fig.~\ref{fig6}), which requires a decreasing $D_{\mathrm{Ca}^{2+}}^{\mathrm{p}}$ to maintain agreement with the experimental conductance (Fig.~\ref{fig7}A). 
Similarly, the fitted Cl$^{-}$ diffusion coefficient increases because, with the lessened charge inversion, its concentration in the double layer diminishes (blue line in top panel of Fig.~\ref{fig9}).

Although the diffusion coefficient of K$^{+}$ remains nearly constant with increasing $r_0$ (Fig.~\ref{fig7}A), its conductance increases because the average K$^{+}$ concentration inside the pore rises (Fig.~\ref{fig9}, middle panel). 
This reflects the reduced electrostatic repulsion from the pore wall as charge inversion weakens, leading to an increase in $G_{\mathrm{K}^{+}}'$ (Fig.~\ref{fig6}).

\subsection{Explicit-oxygen models}

The results obtained with the fixed point-charge model indicate that the DCA between the ionic charge and the pore charge plays a central role in determining transport behavior. 
A natural next question is whether the fixed nature of the pore charges itself is important. 
Specifically, do the results change if the pore charges are represented by a more realistic, explicit particle-based model of the oxygen atoms, as described in Sec.~\ref{sec:pore_charges}?

Figure~\ref{fig5} already suggests that the detailed representation of the surface groups plays a secondary role and that the DCA is the key parameter. 
The results obtained using the two-oxygen model, shown by small open squares, practically coincide with those of the fixed-charge model at $r_0=0.14$ nm (i.e., when the distance of the fixed point charge from the wall equals the radius of the oxygen atoms, $R_{\mathrm{f}}=0.14$ nm). 
In this case, the DCA is identical in the two models.

\begin{figure}[t!]  
	\centering
	\includegraphics[width=0.9\linewidth]{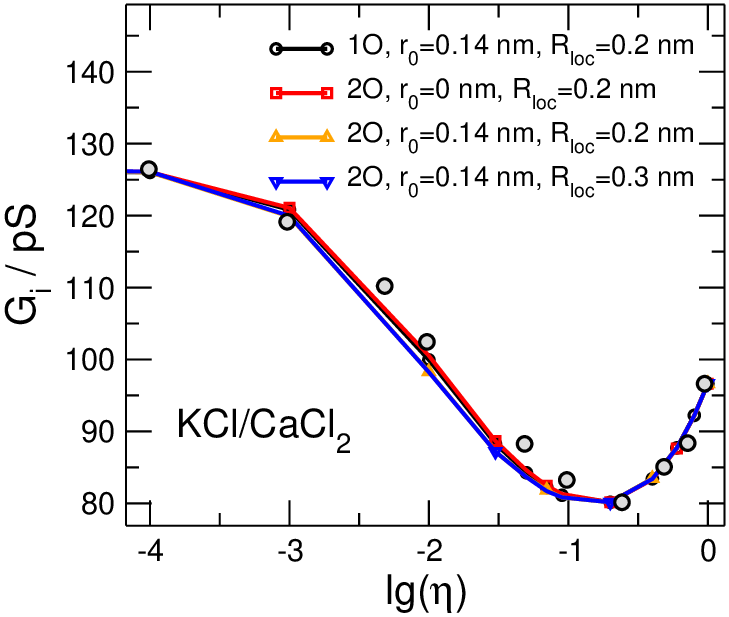}
	\caption{\small Total conductance as a function of Ca$^{2+}$ mole fraction, $\eta$, for different models of the carboxyl groups. Various curves refer to results obtained for modeling the carboxyl groups with one or two oxygen atoms (charged hard spheres with radius $R_\mathrm{f}=0.14$ nm) carrying $-e$ total charge. Results are shown for different values of the $r_0$ and $R_{\mathrm{loc}}$ parameters. The large gray circles denote experimental results~\cite{gillespie_bj_2008_nanopore}. Total concentration is fixed at $0.1$ M.}
	\label{fig10}
\end{figure}

To further support the notion that the DCA serves as an effective control parameter in this reduced model, Fig.~\ref{fig10} presents the total conductance $G(\eta)$ for the four different oxygen models illustrated in Fig.~\ref{fig3}. 
The four curves are virtually indistinguishable, indicating that all models yield the same device-level behavior despite their different microscopic descriptions.

\begin{figure}[t!]  
	\centering
	\includegraphics[width=0.9\linewidth]{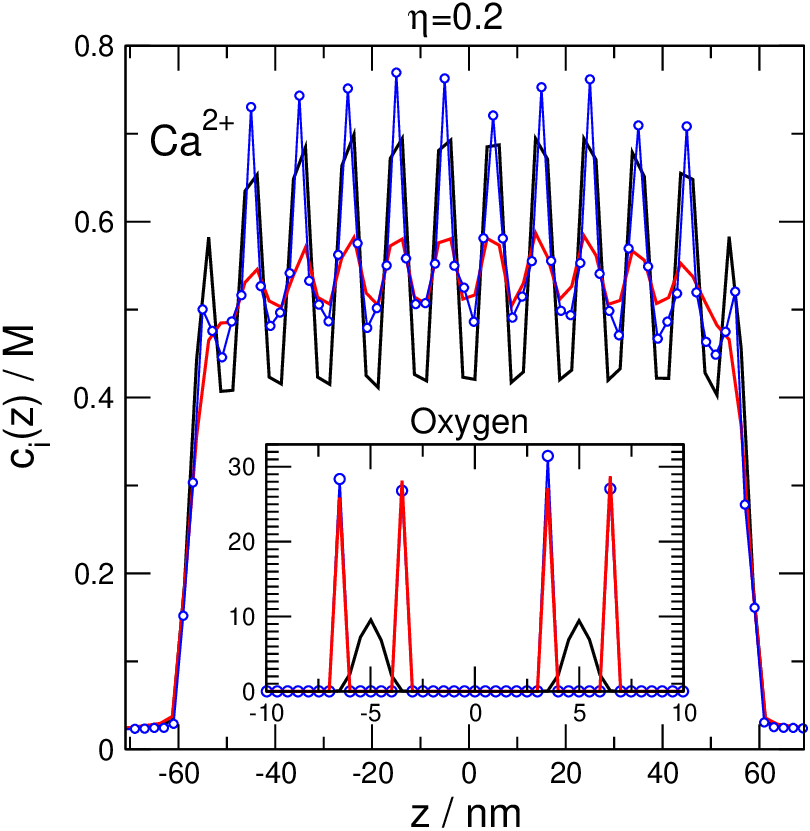}
	\caption{\small Axial concentration profiles of Ca$^{2+}$ (main panel) and  oxygen atoms (inset) for different models of the carboxyl groups in the framework of modeling the oxygen atoms as one or two charged hard spheres with radius $R_\mathrm{f}=0.14$ nm carrying $-e$ total charge. Results are shown for different values of the $r_0$ and $R_{\mathrm{loc}}$ parameters. The black, red, and blue colors have the same meaning as in Fig.~\ref{fig10}. Profiles are shown for CaCl$_{2}$ mole fraction $\eta = 0.2$.}
	\label{fig11}
\end{figure}

This result is particularly striking given that the models exhibit markedly different local ion distributions for Ca$^{2+}$ ions and oxygens. 
As shown in Fig.~\ref{fig11}, the axial concentration profiles of Ca$^{2+}$ display peaks of varying height and depletion zones of different depth across the four models. 
The distributions of the oxygen atoms also differ, especially when comparing models with one versus two oxygen atoms (Fig.~\ref{fig11} inset).

The apparent paradox---distinct Ca$^{2+}$ concentration profiles yielding identical conductance curves---is resolved by noting that the concentration profiles of K$^{+}$ and Cl$^{-}$ ions are essentially identical across all models (results not shown). 
These ions remain sufficiently distant from the oxygen atoms that their distributions, and thus their $G_i'$ values, are unaffected by the details of the surface-charge representation. 
Consequently, fitting the model to the three experimental data points effectively accounts for the Ca$^{2+}$ contribution to the total conductance by adjusting $D_{\mathrm{Ca}^{2+}}^{\mathrm{p}}$, leading to the same macroscopic $G(\eta)$ behavior in all cases.

\subsection{NaCl/CaCl\textsubscript{2} and LiCl/CaCl\textsubscript{2} mixtures}

The study by Gillespie et al.~\cite{gillespie_bj_2008_nanopore} also reports AMFE measurements for mixtures of LiCl/CaCl$_2$ and NaCl/CaCl$_2$, in addition to KCl/CaCl$_2$. 
Because the $\eta=1$ limit (pure CaCl$_2$) is identical for all mixtures, there is no modifying the pore diffusion coefficients $D_{\mathrm{Ca}^{2+}}^{\mathrm{p}}$ and $D_{\mathrm{Cl}^{-}}^{\mathrm{p}}$ obtained from the KCl/CaCl$_2$ system. 
Accordingly, only the diffusion coefficients of the monovalent cations, $D_{\mathrm{Na}^{+}}^{\mathrm{p}}$ and $D_{\mathrm{Li}^{+}}^{\mathrm{p}}$, are fitted to the experimental data at $\eta=0$. 

\begin{figure}[t!]  
	\centering
	\includegraphics[width=0.9\linewidth]{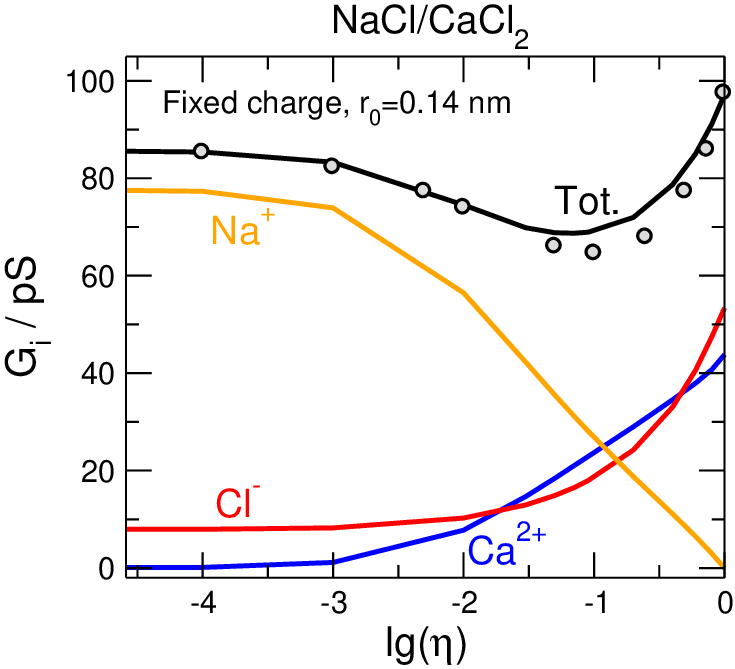}
	\caption{\small Conductances of ionic species (Ca$^{2+}$, Na$^{+}$, and Cl$^{-}$) and total conductance as functions of Ca$^{2+}$ mole fraction, $\eta$, for the fixed-charge model with the point charge being ``behind the wall'' at distance $r_0=0.14$ nm from the wall. The large gray circles denote experimental results~\cite{gillespie_bj_2008_nanopore}. Total concentration is fixed at $0.1$ M.}
	\label{fig12}
\end{figure}

\begin{figure}[t!]  	
	\centering
	\includegraphics[width=0.9\linewidth]{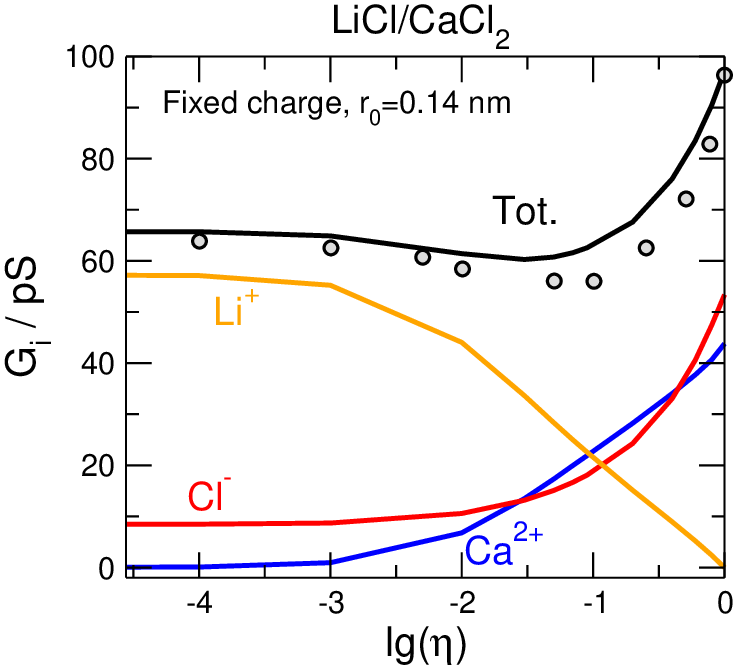}
	\caption{\small Conductances of ionic species (Ca$^{2+}$, Li$^{+}$, and Cl$^{-}$) and total conductance as functions of Ca$^{2+}$ mole fraction, $\eta$, for the fixed-charge model with the point charge being ``behind the wall'' at distance $r_0=0.14$ nm from the wall. The large gray circles denote experimental results~\cite{gillespie_bj_2008_nanopore}. Total concentration is fixed at $0.1$ M.}
	\label{fig13}
\end{figure}

Figures~\ref{fig12} and~\ref{fig13} show the conductance curves for NaCl/CaCl$_2$ and LiCl/CaCl$_2$, respectively, obtained using the fixed-charge (behind-the-wall) model with $r_0=0.14$ nm. 
The agreement between the computed and experimental $G(\eta)$ curves is reasonable for NaCl and noticeably poorer for LiCl. 
In both cases, the predicted minima around $\eta\approx0.2$ are too shallow in the simulations, indicating that Ca$^{2+}$ competes less effectively with Na$^{+}$ and Li$^{+}$ for occupancy near the pore charges than it does with K$^{+}$.

This behavior can be traced to the use of Pauling radii for the ions, which are relatively small for Na$^{+}$ and Li$^{+}$ (Table~\ref{tab}). 
Within the charged-hard-sphere framework, these small radii allow Na$^{+}$ and Li$^{+}$ to approach the pore charges closely and thus compete efficiently with Ca$^{2+}$ (Fig.~7 of Ref.\cite{boda_jgp_2009}). 
In reality, however, these smaller monovalent ions retain more tightly bound hydration shells than K$^{+}$, which reduces their ability to displace Ca$^{2+}$ from the charged surface. 
This effect could be incorporated by assigning larger effective (hydrated) radii to Na$^{+}$ and Li$^{+}$, but, given the complicated nature of ion hydration, this refinement is not pursued in the present work.

\section{Conclusions}

In this work, we investigated the AMFE in a negatively charged PET nanopore using a reduced NP+LEMC modeling framework, with a particular focus on how the microscopic representation of surface carboxyl groups affects charge inversion and ionic transport. 
By systematically varying the model of the pore charges (from fixed point charges to mobile explicit oxygen atoms) we identified that the level of charge inversion is modulated via two key parameters, the DCA between ionic charges and pore charges and the spacing of the pore charge grid (e.g., localization) to control device-level transport behavior.

Our results demonstrate that charge inversion, driven by strong local electrostatic attraction, leads to pronounced Ca$^{2+}$ adsorption and deep axial depletion zones in the concentration profiles.
These features partially determine the current through the $c_{i}(\mathbf{r})\nabla \mu_{i}(\mathbf{r})$ part of the NP equation.
The concentration profiles are obtained from NP+LEMC calculations based on a specific model for the pore charges. 
The current is also tuned by the diffusion coefficients in the pore, $D_{i}^{\mathrm{p}}$, that are obtained from fitting to experimental data.
This approach reflects our goals of reproducing, interpreting, and understanding experimental results with relatively simple reduced models.

Changing the model of the pore charge (particularly, DCA and grid spacing) modulates the strength of charge inversion that, in turn, changes the concentration profiles and the $D_{i}^{\mathrm{p}}$-independent $G_{i}'$.
Different diffusion constants in the pore, therefore, define the relative importance of each ion species to the total pore current.
Our results show that the physically sensible model parameters $Q=-e$, $\Delta x=1$ nm, and $\mathrm{DCA}=R_{\mathrm{ion}}+R_{\mathrm{f}}$ lead to a balance of Ca$^{2+}$ and Cl$^{-}$ conductances that is in agreement with experiments~\cite{he_jacs_2009} and MD simulations~\cite{shabbir_jcp_2026}.

A central finding of this study is that reduced models with limited structural detail can reproduce experimental observables.
As soon as the important features (localization and DCA) are appropriately chosen, the model can reproduce not only the whole AMFE curve characterizing preferential selectivity of Ca$^{2+}$ over K$^{+}$, but also provides results for Ca$^{2+}$ versus K$^{+}$ selectivity that are in agreement with other experiments~\cite{he_jacs_2009} and MD simulations~\cite{shabbir_jcp_2026}.

Finally, our analysis of NaCl/CaCl$_{2}$ and LiCl/CaCl$_{2}$ mixtures indicates that the reduced charged-hard-sphere model overestimates the ability of small monovalent ions to compete with Ca$^{2+}$, likely due to our neglecting ion-specific hydration effects,  which is a known shortcoming~\cite{vincze_jcp_2010}. 
While these effects were taken into account for K$^{+}$ by the fitted diffusion coefficients, the model for Na$^{+}$ and Li$^{+}$ ``inherited'' the value of $D_{\mathrm{Cl}^{-}}$ from the CaCl$_{2}$/KCl calculations.
To bring the simulated $G(\eta)$ curves into agreement with experimental AMFE curves, additional information is needed either in the form of a refitted $D_{\mathrm{Cl}^{-}}$ value, or in the form of fitted hydrated ionic radii.

Overall, this work clarifies the physical mechanisms underlying the AMFE in wide nanopores and provides a clear modeling guideline: accurate representation of the effective ion-surface distance is more important than the detailed atomistic structure of the surface groups. 
This insight strengthens the case for reduced models as powerful tools for interpreting ionic selectivity in nanofluidic systems.

\section*{Acknowledgements}
\label{sec:ack}

This work has been implemented by the National Multidisciplinary Laboratory for Climate Change (RRF-2.3.1-21-2022-00014) project within the framework of Hungary's National Recovery and Resilience Plan supported by the Recovery and Resilience Facility of the European Union.
We gratefully acknowledge  the financial support of the National Research, Development, and Innovation Office -- NKFIH K137720 and the TKP2021-NKTA-21.
We are grateful to Zolt\'an Hat\'o for inspiring discussions, Salman Shabbir for the MD data, and Pavel Apel for discussion on Fig.~1A.

\bibliography{nanopore,book,own}

@book{haynes2016crc,
	title        = {CRC Handbook of Chemistry and Physics},
	editor       = {Haynes, William M.},
	edition      = {97th},
	publisher    = {CRC Press/Taylor \& Francis},
	address      = {Boca Raton, FL},
	year         = {2016},
	isbn         = {978-1498754286}
}

@Book{Hille,
  Title                    = {{Ion channels of excitable membranes}},
  Author                   = {Hille, B.},
  Publisher                = {Sinauer Associates},
  Year                     = {2001},

  Address                  = {Sunderland},
  Edition                  = {3rd}
}

@Book{pauling,
  Title                    = {{The Nature of the Chemical Bond}},
  Author                   = {Pauling, L.},
  Publisher                = {Cornell Universiry Press},
  Year                     = {1960},

  Address                  = {Ithaca, NY},
  Edition                  = {3rd}
}

@article{YuApel2012,
	title = {Asymmetric ion track nanopores for sensor technology. Reconstruction of pore profile from conductometric measurements},
	volume = {23},
	ISSN = {1361-6528},
	url = {http://dx.doi.org/10.1088/0957-4484/23/22/225503},
	DOI = {10.1088/0957-4484/23/22/225503},
	number = {22},
	journal = {Nanotechnology},
	publisher = {IOP Publishing},
	author = {Yu Apel, P. and Blonskaya, I. V. and Orelovitch, O. L. and Sartowska, B. A. and Spohr, R.},
	year = {2012},
	pages = {225503}
}

@ARTICLE{Yu-et-al,
  author = {Yu, H. and Noskov, S. Y. and Roux, B.},
  title = {Two mechanisms of ion selectivity in protein binding sites},
  journal = {Proc. Natl. Acad. Sci. U. S. A.},
  year = {2010},
  volume = {107(47)},
  pages = {20329--20334}
}

@ARTICLE{Finnerty,
  author = {Finnerty, J. J. and Eisenberg, R. and Carloni, P.},
  title = {Localizing the Charged Side Chains of Ion Channels within the Crowded Charge Models},
  journal = {J. Chem. Theory Comput.},
  year = {2012},
  volume = {9(1)},
  pages = {766--773}
}

@Article{gillespie_bj_2008_energetics,
  Title                    = {Energetics of Divalent Selectivity in a Calcium Channel: {The} {Ryanodine Receptor} Case Study},
  Author                   = {D. Gillespie},
  Journal                  = {Biophys. J.},
  Year                     = {2008},
  Number                   = {4},
  Pages                    = {1169--1184},
  Volume                   = {94},

  Doi                      = {10.1529/biophysj.107.116798},
  Publisher                = {Elsevier {BV}},
  Url                      = {https://doi.org/10.1529%2Fbiophysj.107.116798}
}

@Article{gillespie_giri_bj_2009,
  author  = {Gillespie, D. and Giri, J. and Fill, M.},
  title   = {{Reinterpreting the anomalous mole fraction effect: {The} {Ryanodine} receptor case study}},
  journal = {Biophys. J.},
  year    = {2009},
  volume  = {97},
  number  = {8},
  pages   = {2212--2221},
  doi     = {10.1016/j.bpj.2009.08.009},
  url     = {https://doi.org/10.1016/j.bpj.2009.08.009},
}

@Article{gillespie_jpcb_2005,
  author  = {Gillespie, D. and Xu, L. and Wang, Y. and Meissner, G.},
  title   = {{({De})constructing the Ryanodine Receptor: {Modeling} Ion Permeation and Selectivity of the Calcium Release Channel}},
  journal = {J. Phys. Chem. B},
  year    = {2005},
  volume  = {109},
  number  = {32},
  pages   = {15598--15610},
  doi     = {10.1021/jp052471j},
  url     = {https://doi.org/10.1021/jp052471j},
}

@Article{almers_jp_1984b,
	author = {W. Almers and E. W. McCleskey},
	title = {Non-selective conductance in calcium channels of frog muscle: {c}alcium selectivity in a single-file pore},
	journal = {J. Physiol.},
	year = {1984},
	volume = {353},
	pages = {585--608}
}

@Article{almers_jp_1984a,
  author    = {W. Almers and E. W. McCleskey and P. T. Palade},
  title     = {A non-selective cation conductance in frog muscle membrane blocked by micromolar external calcium ions.},
  journal   = {J. Physiol.},
  year      = {1984},
  volume    = {353},
  number    = {1},
  pages     = {565--583},
  doi       = {10.1113/jphysiol.1984.sp015351},
  publisher = {Wiley},
  url       = {https://doi.org/10.1113%2Fjphysiol.1984.sp015351},
}

@Article{shabbir_jcp_2026,
  author  = {S. Shabbir and D. Boda and Z. Hat\'o},
  title   = {Interplay of ion availability and mobility in the loss of cation selectivity for {CaCl}\textsubscript{2} in negatively charged nanopores: molecular dynamics using scaled-charge models.},
  note = {{https://arxiv.org/abs/2602.10600}},
  year    = {2026}
}

@article{fabian_jml_2022,
	title = {Calcium versus potassium selectivity in a nanopore: {The} effect of charge inversion at localized pore charges},
	journal = {J. Mol. Liq.},
	volume = {368},
	pages = {120715},
	year = {2022},
	issn = {0167-7322},
	doi = {https://doi.org/10.1016/j.molliq.2022.120715},
	url = {https://www.sciencedirect.com/science/article/pii/S0167732222022541},
	author = {H. F\'abi\'an and Zs. Sarkadi and M. Valisk\'o and D. Gillespie and D. Boda}
}

@Article{valisko_jcp_2014,
author = {M. Valisk\'o and D. Boda},
title = {The effect of concentration- and temperature-dependent dielectric constant on the activity coefficient of {NaCl} electrolyte solutions},
journal = {J. Chem. Phys.},
year = {2014},
volume = {140},
number = {23},
pages = {234508}}

@Article{valisko_jpcb_2015,
  author = {M. Valisk\'{o} and D. Boda},
  title = {Unraveling the behavior of the individual ionic activity coefficients on the basis of the balance of ion-ion and ion-water interactions},
  journal = {J. Phys. Chem. B},
  volume = {119},
  number = {4},
  pages = {1546--1557},
  year = {2015}
}

@Article{valisko_mp_2017,
  author = {M. Valisk\'{o} and D. Boda},
  title = {Activity coefficients of individual ions in LaCl$_{3}$ from the II+IW theory},
  journal = {Mol. Phys.},
  volume = {115},
  number = {9--12},
  pages = {1245-1252},
  year = {2017}
}

@article{boda_entropy_2020,
	doi = {10.3390/e22111259},
	url = {https://doi.org/10.3390%2Fe22111259},
	year = {2020},
	publisher = {{MDPI} {AG}},
	volume = {22},
	number = {11},
	pages = {1259},
	author = {D. Boda and M. Valisk{\'{o}} and D. Gillespie},
	title = {Modeling the Device Behavior of Biological and Synthetic Nanopores with Reduced Models},
	journal = {Entropy}
}

@Article{valisko_jcp_2019,
  author  = {M. Valisk\'o and B. Matejczyk and Z. Hat\'o and T. Krist\'of and E. M\'adai and D. Fertig and D. Gillespie and D. Boda},
  title   = {Multiscale analysis of the effect of surface charge pattern on a nanopore's rectification and selectivity properties: from all-atom model to Poisson-Nernst-Planck},
  journal = {J. Chem. Phys.},
  year    = {2019},
  volume  = {150},
  number  = {14},
  pages   = {144703},
  doi     = {10.1063/1.5091789},
  url     = {https://doi.org/10.1063%2F1.5091789},
}

@InCollection{boda_arcc_2014,
  author    = {D. Boda},
  booktitle = {Ann. Rep. Comp. Chem.},
  publisher = {Elsevier},
  year      = {2014},
  editor    = {R. A. Wheeler},
  volume    = {10},
  chapter   = {5 {Monte Carlo} Simulation of Electrolyte Solutions in Biology: {In} and Out of Equilibrium},
  pages     = {127--163},
  doi       = {10.1016/b978-0-444-63378-1.00005-7},
  url       = {https://doi.org/10.1016%2Fb978-0-444-63378-1.00005-7},
}

@Article{boda_jpcb_2000,
  author  = {D. Boda and D. D. Busath and D. Henderson and S. Soko{\l}owski},
  title   = {Monte {Carlo} simulations of the mechanism for channel selectivity: {The} competition between volume exclusion and charge neutrality},
  journal = {J. Phys. Chem. B},
  year    = {2000},
  volume  = {104},
  number  = {37},
  pages   = {8903--8910},
  doi     = {10.1021/jp0019658},
  url     = {https://doi.org/10.1021%2Fjp0019658},
}

@Article{boda_jctc_2012,
  author  = {D. Boda and D. Gillespie},
  title   = {Steady state electrodiffusion from the {Nernst-Planck} equation coupled to {Local Equilibrium Monte Carlo} simulations},
  journal = {J. Chem. Theor. Comput.},
  year    = {2012},
  volume  = {8},
  number  = {3},
  pages   = {824--829},
  doi     = {10.1021/ct2007988},
  url     = {https://doi.org/10.1021%2Fct2007988},
}

@Article{boda_jcp_2011_analyze,
  author  = {D. Boda and J. Giri and D. Henderson and B. Eisenberg and D. Gillespie},
  title   = {Analyzing the components of the free energy landscape in a calcium selective ion channel by {Widom}'s particle insertion method},
  journal = {J. Chem. Phys.},
  year    = {2011},
  volume  = {134},
  number  = {5},
  pages   = {055102},
  doi     = {10.1063/1.3532937},
  url     = {https://doi.org/10.1063/1.3532937},
}

@Article{boda_mp_2002,
  author  = {D. Boda and D. Henderson and D. D. Busath},
  title   = {Monte {Carlo} study of the selectivity of calcium channels: improved geometrical model},
  journal = {Mol. Phys.},
  year    = {2002},
  volume  = {100},
  number  = {14},
  pages   = {2361--2368},
  doi     = {10.1080/00268970210125304},
  url     = {https://doi.org/10.1080%2F00268970210125304},
}

@Article{boda_jcp_2013_solvation,
  author  = {D. Boda and D. Henderson and D. Gillespie},
  title   = {The role of solvation in the binding selectivity of the {L}-type calcium channel},
  journal = {J. Chem. Phys.},
  year    = {2013},
  volume  = {139},
  number  = {5},
  pages   = {055103},
  doi     = {10.1063/1.4817205},
  url     = {https://doi.org/10.1063%2F1.4817205},
}

@Article{boda_jcp_2006,
  author  = {D. Boda and M. Valisk{\'o} and B. Eisenberg and W. Nonner and D. Henderson and D. Gillespie},
  title   = {The effect of protein dielectric coefficient on the ionic selectivity of a calcium channel},
  journal = {J. Chem. Phys.},
  year    = {2006},
  volume  = {125},
  number  = {3},
  pages   = {034901},
  doi     = {10.1063/1.2212423},
  url     = {https://doi.org/10.1063%2F1.2212423},
}

@Article{boda_jgp_2009,
  author  = {D. Boda and M. Valisk{\'o} and D. Henderson and B. Eisenberg and D. Gillespie and W. Nonner},
  title   = {Ion selectivity in {L-type} calcium channels by electrostatics and hard-core repulsion},
  journal = {J. Gen. Physiol.},
  year    = {2009},
  volume  = {133},
  number  = {5},
  pages   = {497--509},
  doi     = {10.1085/jgp.200910211},
  url     = {https://doi.org/10.1085%2Fjgp.200910211},
}

@Article{gillespie_bj_2008_ca,
  author  = {D. Gillespie and D. Boda},
  title   = {The anomalous mole fraction effect in calcium channels: {A} measure of preferential selectivity},
  journal = {Biophys. J.},
  year    = {2008},
  volume  = {95},
  number  = {6},
  pages   = {2658--2672},
  doi     = {10.1529/biophysj.107.127977},
  url     = {https://doi.org/10.1529%2Fbiophysj.107.127977},
}

@Article{gillespie_bj_2008_nanopore,
  author  = {D. Gillespie and D. Boda and Y. He and P. Apel and Z.S. Siwy},
  title   = {Synthetic nanopores as a test case for ion channel theories: {The} anomalous mole fraction effect without single filing},
  journal = {Biophys. J.},
  year    = {2008},
  volume  = {95},
  number  = {2},
  pages   = {609--619},
  doi     = {10.1529/biophysj.107.127985},
  url     = {https://doi.org/10.1529%2Fbiophysj.107.127985},
}

@Article{he_jacs_2009,
  author  = {Y. He and D. Gillespie and D. Boda and I. Vlassiouk and R. S. Eisenberg and Z. S. Siwy},
  title   = {Tuning transport properties of nanofluidic devices with local charge inversion},
  journal = {JACS},
  year    = {2009},
  volume  = {131},
  number  = {14},
  pages   = {5194--5202},
  doi     = {10.1021/ja808717u},
  url     = {https://doi.org/10.1021%2Fja808717u},
}

@Article{malasics_bba_2010_trivalent,
  author  = {M. Malasics and D. Boda and M. Valisk{\'o} and D. Henderson and D. Gillespie},
  title   = {Simulations of calcium channel block by trivalent ions: {Gd$^{3+}$} competes with permeant ions for the selectivity filter},
  journal = {Biochim. et Biophys. Acta - Biomembranes},
  year    = {2010},
  volume  = {1798},
  number  = {11},
  pages   = {2013--2021},
  doi     = {10.1016/j.bbamem.2010.08.001},
  url     = {https://doi.org/10.1016%2Fj.bbamem.2010.08.001},
}

@Article{vincze_jcp_2010,
  author  = {J. Vincze and M. Valisk{\'o} and D. Boda},
  title   = {The nonmonotonic concentration dependence of the mean activity coefficient of electrolytes is a result of a balance between solvation and ion-ion correlations},
  journal = {J. Chem. Phys.},
  year    = {2010},
  volume  = {133},
  number  = {15},
  pages   = {154507},
}
\bibliographystyle{unsrt}

\end{document}